\documentstyle[preprint,aps,eqsecnum]{revtex}
\begin{document}
\draft
\tighten

\title{\Large\bf Nonperturbative Description of Deep Inelastic 
	Structure Functions in Light-Front QCD}

\author{{\bf A. Harindranath$^a$, Rajen Kundu$^{a}$, and Wei-Min Zhang$^b$} \\
$^a$Saha Institute of Nuclear Physics, 1/AF, Bidhan Nagar, 
	Calcutta 700064 India \\
$^b$Institute of Physics, Academia Sinica, Taipei, Taiwan 11529, ROC}

\date{Feb. 5, 1998}

\maketitle

\begin{abstract}
In this paper, we explore the deep inelastic 
structure functions of hadrons nonperturbatively
in an inverse power expansion 
of the light-front energy of the probe in the framework of 
light-front QCD.  We arrive at the general
expressions for various structure functions as the Fourier 
transform of matrix elements of different components of bilocal 
vector and axial vector currents on the light-front
in a straightforward manner. 
The complexities of the structure functions are 
mainly carried by the multi-parton wave functions of the hadrons, while, 
the bilocal currents have a dynamically dependent yet simple structure 
on the light-front in this description. 
We also present a novel analysis of the power corrections 
based on light-front power counting which resolves some ambiguities 
of the conventional twist analysis in deep inelastic processes.
Further, the factorization theorem and the scale evolution of the 
structure functions are presented 
in this formalism by using 
old-fashioned light-front time-ordered perturbation theory with 
multi-parton wave functions. Nonperturbative QCD 
dynamics underlying the structure functions can be explored
in the same framework.
Once the nonperturbative multi-parton wave functions are known
from low-energy light-front QCD, a complete description of deep 
inelastic structure functions can be realized.
\end{abstract}

\vspace{0.5in}


%\vspace{0.5in}

%%%%%%%%%%%%%%%%%%%%%%%%%%%%%%%%%%%%%%%%%%%%%%%%%%%%%%%%%%%%%%%%%
\section{introduction}
%%%%%%%%%%%%%%%%%%%%%%%%%%%%%%%%%%%%%%%%%%%%%%%%%%%%%%%%%%%%%%%%%
With the recent and planned experiments on polarized and unpolarized
structure functions, the field of deep inelastic scattering  has 
entered a new era. New experiments are beginning to provide invaluable 
information on the so-called ``higher twist" (power suppressed) 
contributions to deep inelastic cross sections, a theoretical 
understanding of which requires nonperturbative information on the 
structure of hadrons. To unravel this structure, there is an urgent 
need to develop nonperturbative theoretical tools which are preferably 
based on physical intuition and which at the same time employs well-defined 
field theoretical calculational procedures. Towards this goal, in this 
work, we propose a new method of calculation of structure functions
combining the coordinate space approach based on light-front current 
algebra techniques and the momentum space approach based on Fock space 
expansion methods in light-front theory in a Hamiltonian QCD framework.   

To get an intuitive picture of deep inelastic scattering in field theory, 
it is extremely helpful to keep close contact with parton ideas. However,
partons were originally introduced as collinear, massless, on-mass shell
objects. In reality, the QCD governed interacting partons should not be 
collinear and massless. The question, then, is can
one generalize this concept and introduce {\it field theoretic partons} as
non-collinear and massive (in the case of quarks) but still on-mass shell
objects in interacting field theory? Nonperturbative light-front Hamiltonian 
description of composite systems which utilizes many body wave functions 
for the constituents allows us to precisely achieve this goal.  

Now, the problem is how to introduce many body (or multi-parton) wave 
functions in the description of deep inelastic structure functions. An
attractive possible avenue is provided by the BJL (high energy) expansion 
of scattering amplitudes together with the use of light-front current
algebra. This is essentially {\it a nonperturbative approach} where the 
expansion parameter is the inverse of the light-front energy of the 
probe (in the present case, the virtual photon). In this approach one 
can arrive at expressions for various structure functions as the Fourier 
Transform (in the light-front longitudinal direction) of matrix elements 
of different components of bilocal vector and axial vector currents
in {\it light-front field theory}.  

In the standard approach to deep inelastic scattering based on Wilson's
Operator Product Expansion (OPE) method which is more mathematical, 
one considers the problem of renormalizing
composite operators. In contrast, in the current algebra 
approach, products of field operators are evaluated at equal {\it 
light-front} time and have the same form as their free field theory 
counterparts. Hence in this approach, most of the complexities appear to 
be carried by the hadronic states. We demonstrate that an  expansion of 
the state allows us to exhibit this complexity manifestly in terms of 
the multi-parton wave functions, where the constituents are on-mass
shell objects with non-vanishing transverse momenta. The nonperturbative
nature of the matrix elements relevant for various structure functions are
thus directly translated into the language of multi-parton wave functions. 
The structure functions, then, can be conveniently evaluated once the
nonperturbative wave functions renormalized at scale $Q$ are known. At 
present major efforts are underway to evaluate these wave functions. 

With the advent of QCD, the current algebra approach which originally 
lead to the prediction of scaling without introducing the concept of 
partons was altogether abandoned in favor of the OPE method, primarily 
because it was known that the canonical manipulations that lead to 
many of the current algebra predictions are invalidated in perturbation 
theory which gives rise to logarithmic corrections. The only exceptions 
were  sum rules protected by conservation laws. Obviously, what was 
missing in the current algebra approach was the realization that the 
{\it matrix elements} relevant for deep inelastic scattering are those 
renormalized at the physical scale $Q$. Thus one of our  major problem
is whether we can consistently carry out renormalization procedure. 
In a subsequent paper \cite{deep2}, we shall demonstrate that in the 
case of leading order structure functions the scaling violations of 
perturbation theory can be successfully addressed by replacing the 
hadron target by a dressed parton target in the matrix element and 
carrying out a well-defined perturbative expansion which closely follows 
the techniques of (light-front) time-ordered perturbation theory 
\cite{brodsky80,zhang93,Hari96}. Here, we will show that we can address 
the issues 
of factorization and scale evolution in the case of a hadron target by 
separating the soft and hard parts of the multi-parton wave functions. 
Furthermore, the nonperturbative contributions to deep inelastic 
structure functions can be addressed {\it within the same framework} by 
incorporating the newly developed light-front renormalization group 
approach to nonperturbative QCD \cite{Wilson94,zhang97,Perry97}.  Therefore,
a unified light-front description of the perturbative and nonperturbative
QCD underlying the deep inelastic structure functions can be realized.

The paper is organized as follows. In Sec. II, a brief overview of deep
inelastic structure functions is presented. In Sec. III, we discuss 
within light-front QCD how the deep inelastic structure functions are
nonperturbatively related to the bilocal operators in 
light-front current algebra 
through an inverse power expansion of light-front energy of the 
probe. In Sec. IV, we determine the structure functions in 
terms of the matrix elements of the light-front 
bilocal vector and axial vector currents separated only in the 
longitudinal direction. The operator structures involved in the 
structure functions are explored and the complexities of structure
functions are analyzed in Sec. V, where a scheme to evaluate the hard
and soft contributions to deep inelastic structure functions in the 
light-front time-ordering Hamiltonian formalism is proposed. Meanwhile, 
the concept of twist is examined and redefined on the light-front, 
which removes some ambiguities in previous works. In Sec. VI, we 
re-derive sum rules the structure functions obey and discuss their 
physical implications. 
Finally a summary is given in Sec. VII.
 
%%%%%%%%%%%%%%%%%%%%%%%%%%%%%%%%%%%%%%%%%%%%%%%%%%%%%%%%%%%%%%%%
\section{A brief overview on deep inelastic structure functions}
%%%%%%%%%%%%%%%%%%%%%%%%%%%%%%%%%%%%%%%%%%%%%%%%%%%%%%%%%%%%%%%%     
We begin with a brief review of the basic ingredients of  
lepton-nucleon deep inelastic scattering (DIS):
\begin{equation}
	e(k) + h(P) \longrightarrow e(k-q)  + X(P+q) \, .
\end{equation}
The cross section for the above scattering process is given by 
\begin{equation}
	{d \sigma \over d\Omega dE'} = {1\over 2M}{\alpha^2 \over q^4}
		{E'\over E} L_{\mu \nu} W^{\mu \nu} \, ,
\end{equation}
where $E$ ($E'$) is the energy of the incoming (outgoing) lepton,
$L_{\mu \nu}$ is the leptonic matrix element,
\begin{eqnarray}
	L_{\mu \nu} &=& {1\over 2}\sum_{s'}[\overline{u}(k,s)\gamma_\mu 
		u(k',s') \overline{u}(k',s')\gamma_\nu u(k,s)] \nonumber \\
	&=& 2(k'_\mu k_\nu + k'_\nu k_\mu) - 2 g_{\mu \nu} k \cdot k'
		- 2i \epsilon_{\mu \nu \rho \sigma} q^\rho s^\sigma \, ,
\end{eqnarray}
and $W^{\mu \nu}$ the hadronic tensor which contains all the hadronic 
dynamics involved in DIS process, 
\begin{equation}
	W^{\mu \nu} = {1\over 4\pi} \int d^4 \xi~ e^{iq \cdot \xi} 
		\langle PS |[J^\mu(\xi), J^\nu(0)]|PS \rangle  \, ,
\end{equation}
where $P$ and $S$ are the target four-momentum and polarization 
vector respectively ($P^2=M^2, S^2=-M^2, S\cdot P=0$), $q$ is 
the virtual-photon four momentum, and $J^\mu(x)=\sum_\alpha 
e_\alpha \overline{\psi}_\alpha(x) \gamma^\mu \psi_\alpha(x)$ the 
electromagnetic current with quark field $\psi_\alpha (x)$ carrying
the flavor index $\alpha$ and the charge $e_\alpha$. 

The above hadronic tensor can be decomposed into independent
Lorentz invariant functions:
\begin{eqnarray}
	W^{\mu \nu} &=&\Big(-g^{\mu \nu} + {q^\mu q^\nu 
		\over q^2} \Big) W_1(x,Q^2) + \Big(P^\mu - {\nu 
		\over q^2} q^\mu\Big)\Big(P^\nu -{\nu \over q^2} 
		q^\nu\Big)W_2(x,Q^2) \nonumber \\
	& & - i \epsilon^{\mu \nu \lambda \sigma}q_\lambda \Big[
		S_\sigma W_3(x,Q^2)+ P_\sigma S\cdot q W_4(x,Q^2) 
		\Big] \nonumber \\
	&=& {1\over 2} \Big( g^{\mu \nu} - {q^\mu q^\nu \over q^2} \Big)
		F_L (x,Q^2) + \Big [P^\mu P^\nu - {\nu \over q^2} \Big(
		P^\mu q^\nu + P^\nu q^\mu \Big) + g^{\mu \nu} {\nu^2 
		\over q^2} \Big] {F_2(x,Q^2) \over \nu} \nonumber \\
	& & - i \epsilon^{\mu \nu \lambda \sigma}{q_\lambda \over \nu}
		\Big[ S_{\sigma L} g_1(x,Q^2) + S_{\sigma T} g_T
		(x,Q^2) \Big] \, . \label{wfun}
\end{eqnarray}
The dimensionless functions
\begin{equation}
	F_L(x,Q^2)= 2 \Big [-W_1 + \big [ M^2 -{(P.q)^2 \over q^2}\big ] W_2
\Big ],
\end{equation}
and 
\begin{equation}
 F_2(x,Q^2) = \nu W_2(x,Q^2)
\end{equation}
are the so-called unpolarized structure functions measured from
the unpolarized target and 
\begin{equation}
	g_1(x,Q^2) =\nu \Big[ W_3(x,Q^2) + \nu W_4 (x,Q^2) \Big]~~~,~~~~~	
\end{equation}
and 
\begin{equation}
	g_T(x,Q^2) = g_1(x,Q^2) + g_2(x,Q^2) = \nu W_3(x,Q^2)~~, 
\end{equation}
are the longitudinal and transverse polarized structure functions. While, 
$x={Q^2 \over 2\nu}$ is the famous Bjorken scaling variable, $Q^2=-q^2$  
the momentum transfer carried by the virtual photon and $\nu = P\cdot q$ 
the energy transfer. The longitudinal and transverse polarization vector 
component are given by
\begin{equation}
	S_{\mu L} = S_\mu - S_{\mu T}~~, ~~~~~
	S_{\mu T} = S_\mu - P_\mu {S \cdot q \over \nu} \, .
\end{equation}

These structure functions provide a probe to explore various aspects of 
the intrinsic structure of hadrons. It may be worth noting that in the 
literature, $g_1$ and $g_2$ are usually used to characterize the 
longitudinal and transverse polarized structure functions. However, 
$g_2$ is not really a transverse polarized structure function. 
It also has no clear physical interpretation. Only $g_T$ which can
be directly measured when the target is polarized along the transverse
direction characterizes the full information on
the transverse polarization structure.

The early SLAC experiment discovered that the structure function 
$F_2$ depends only on the Bjorken scaling variable  $x$, and is independent 
of momentum transfer $Q^2$. This discovery lead to  
the parton picture proposed first by Feynman \cite{Feynman72} 
in which the constituents in hadrons can be treated as point-like 
free particles, i.e., the partons. Of course, the picture of 
point-like and non-interacting
partons is certainly over-simplified. As it is well known now, the 
fundamental constituents (i.e., quarks and gluons) inside hadrons 
are indeed strongly interacting with each other and are governed by the 
fundamental QCD theory. Only in the limit of very large $Q^2$, the 
asymptotic freedom feature of QCD makes the quarks and gluons behave 
as point-like and weakly interacting partons. The scale invariance
of $F_2$ is violated for finite values of $Q^2$. For the past twenty-years, 
QCD investigations on deep inelastic structure functions are mainly 
concentrated on the behavior of the
scaling violation of these structure functions, which is believed 
to be the best way to test QCD as the fundamental theory of  the
strong interaction. 

Among these investigations, two methods have dominated the whole 
research topic: the OPE method \cite{ope} and the QCD improved 
(or field theoretic) parton model based on the factorization scheme
\cite{pqcd}. The structure functions have been 
extensively explored either in terms of its moments in the 
language of OPE or in terms of diagrammatic calculation in the 
language of QCD improved parton model.
Both the methods have made great success in understanding the scale
evolution of the structure functions within the perturbative
QCD domain. Although the deep inelastic scatterings provide a 
novel way to separate the physics of perturbative and nonperturbative 
QCD dynamics and allow the exploration of high energy behavior 
of the constituents inside the hadrons, the structure 
functions themselves, however, are still dominated by the 
nonperturbative strong interaction dynamics. A complete 
understanding of hadrons crucially depends on our 
understanding of nonperturbative QCD. The OPE method addresses 
the structure functions in terms of their moments, which naturally
separates the short-distance and long-distance dynamics but it
also makes the nonperturbative dynamics of long-distance physics
more complicated in terms of the moments. On the other hand, the QCD 
improved  parton model was built in the framework of perturbation theory
with the assumption of collinearity, which simplifies the perturbative 
QCD treatment but it is also unclear how to explore nonperturbative QCD 
dynamics because the nonperturbative QCD dynamics is mainly determined 
by the non-collinear motion of the low energy quarks and gluons. 

In the next section we shall use an inverse power expansion of the 
light-front energy $q^-$ of the virtual photon (the extended  
BJL theorem on the light-front) to extract these deep inelastic 
structure functions. This approach was originally proposed to
study DIS sum rules protected by conservation laws in the pre-QCD 
era\cite{Jackiw}. Here, we shall extend this approach to QCD
without recourse to perturbation theory. Therefore it is essentially
a direct nonperturbative QCD description. The resulting structure functions 
are directly expressed in terms of the hadronic matrix elements of 
light-front bilocal vector and axial vector currents, where the
currents have a relatively simple structure although they are dynamically 
dependent. All the hadron dynamics (including both the perturbative
and nonperturbative dynamics) reside in the multi-parton hadronic 
wave functions. This is a description very different from that of the
OPE method and the QCD improved parton model.

%%%%%%%%%%%%%%%%%%%%%%%%%%%%%%%%%%%%%%%%%%%%%%%%%%%%%%%%%%%%%%%%
\section{An expansion in inverse power of light-front energy of
	 the virtual photon}
%%%%%%%%%%%%%%%%%%%%%%%%%%%%%%%%%%%%%%%%%%%%%%%%%%%%%%%%%%%%%%%%     
The inverse power expansion of the virtual photon light-front energy 
$q^-$ is applied to forward scattering amplitudes. Explicitly, as it 
is well known, the hadronic tensor is related to the forward 
virtual-photon hadron Compton scattering amplitude:
\begin{equation}
	W^{\mu \nu} = {1\over 2\pi}{\rm Im} T^{\mu \nu}
\end{equation}
with
\begin{eqnarray}	\label{tfun}	
	T^{\mu \nu} &=& i \int d^4\xi e^{iq\cdot \xi} \langle PS | 
		T(J^\mu(\xi) J^\nu(0)) |PS \rangle  \nonumber \\
	&=&\Big(-g^{\mu \nu} + {q^\mu q^\nu 
		\over q^2} \Big) T_1(x,Q^2) + \Big(p^\mu - {\nu 
		\over q^2} q^\mu\Big)\Big(p^\nu -{\nu \over q^2} 
		q^\nu\Big)T_2(x,Q^2) \nonumber \\
	& & ~~~~~~~~~~~~~ - i \epsilon^{\mu \nu \lambda \sigma}q_\lambda 
		\Big[ S_\sigma T_3(x,Q^2)+ P_\sigma S\cdot q T_4(x,Q^2) 
		\Big] \, .
\end{eqnarray}
Using the optical theorem, we have
\begin{equation}	\label{twr}
	T_i (x,Q^2) = 2 \int_{-\infty}^\infty d{q'}^+
		{W_i (x',Q^2) \over {q'}^+ - q^+} ~, ~~
		i = 1, 2, 3, 4	\, . 
\end{equation}
Then, as we shall see, the structure functions are connected 
with the light-front bilocal currents through the $1/q^-$ expansion 
of $T^{\mu \nu}$.

An expansion of $T^{\mu \nu}$ in terms of  ${1/q^-}$ was originally 
proposed by Jackiw {\it et al.} \cite{Jackiw} based on BJL theorem.  
The general expansion in $1/q^-$ is given by 
\begin{equation} \label{lfcc}
	T^{\mu \nu} = - \sum_{n=0}^\infty \Big({1\over q^-}
		\Big)^{n+1} \int d\xi^- d^2 \xi_\bot e^{iq\cdot \xi}
	  \langle PS | [(i\partial_\xi^-)^nJ^\mu(\xi), J^\nu(0)]_{
		\xi^+=0}| PS\rangle \, ,
\end{equation}
where $q^-=q^0 - q^z$, the light-front energy of the virtual photon,
and $\partial^-=2{\partial \over \partial \xi^+}$ is a light-front
time derivative. The light-front coordinates of the space-time are
defined by 
\begin{equation}
	\xi^\pm = \xi^0 \pm \xi^3~~, ~~~~ \xi_\bot^i = \xi^i ~
		(i = 1, 2) \, .
\end{equation}
The above expansion shows that the time-ordered matrix element can 
be expanded in terms of an infinite series of equal light-front time 
commutators.

For large $Q^2$ and large $\nu$ limits in DIS, theoretically without 
loss of generality we can always select a Lorentz frame 
such that the light-front energy $q^-$ of the virtual photon 
becomes very large. Then, 
only the leading term in the above expansion is dominant, i.e.,
\begin{equation} \label{lfccl}
	T^{\mu \nu} \stackrel{{\rm large}~q^-}{=} - {1\over q^-}
		\int d\xi^- d^2 \xi_\bot e^{iq\cdot \xi}
	  \langle PS | [J^\mu(\xi), J^\nu(0)]_{\xi^+=0}| PS\rangle \, .
\end{equation}
As a result,
the leading contribution to the deep inelastic structure functions 
is determined by the light-front current algebra. The light-front 
current commutator can be computed directly and exactly from QCD 
(where QCD should be quantized on the light-front time surface $\xi^+
=\xi^0+\xi^3=0$ with the light-front gauge $A_a^+=0$ 
\cite{brodsky80,zhang93}). Hence, all the subsequent derivations 
are exact within the light-front QCD and without further assumptions 
or approximations of the collinear and massless partons that were 
used in the previous derivations \cite{ope,pqcd}.

Explicitly, the basic commutation relation on light-front is
\begin{equation}  \label{qkcom}
	\{ \psi_+(x)~, ~\psi_+^\dagger (y) \}_{x^+=y^+}
	 = \Lambda^+ \delta(x^--y^-) \delta^2(x_\bot-y_\bot) \, ,
\end{equation}
which is exact in the full QCD theory \cite{zhang93},
where $\psi_+(x)$ is called the dynamical component of fermion
field on the light-front: 
\begin{equation}
	\psi(x) = \psi_+(x) + \psi_-(x)~~, ~~~~~ \psi_{\pm}(x) = 
		\Lambda^{\pm}\psi(x)~~, ~~~~ \Lambda^{\pm} = 
		{1\over 2}\gamma^0\gamma^\pm  \, .
\end{equation}
The minus component $\psi_-(x)$ is determined from $\psi_+(x)$
as a result of Dirac equation:
\begin{equation}	\label{qkmc}
	\psi_-(x) = {1\over i\partial^+}\Big(i\alpha_\bot \cdot
		D_\bot + \beta m_q \Big) \psi_+(x) \, .
\end{equation}
Here we have already used the light-front gauge $A^+_a=0$, and $D_\bot
=\partial_\bot - ig A_\bot$ is the transverse component of the covariant 
derivative, $\alpha_\bot^i = \gamma^0 \gamma^i$, $\beta=\gamma^0$.

Because of the above special property of quark (or more generally 
fermion) field on the light-front, the light-front current explicitly 
depends on interaction of the theory, which is very different from 
the usual equal-time formulation. In other words, the fundamental
interaction is manifested explicitly in the light-front
current commutators. 

From Eqs.(\ref{qkcom}) and (\ref{qkmc}), we have
\begin{equation}
	\{ \psi_+(x)~, ~\psi^*_-(y) \}_{x^+=y^+} = {\Lambda^+ \over 4i} 
		\epsilon (x^--y^-) \Big[i \alpha_\bot \cdot D^*_\bot -
		\beta m \Big] \delta^2(x_\bot - y_\bot) \, .
\end{equation}
Thus, after a tedious but straightforward calculation, one can find
that
\begin{eqnarray}
	\Big[ J^+(x)~ , ~J^-(y) \Big]_{x^+=y^+} = \sum_\alpha e_\alpha^2
		\Bigg\{ && \partial^+_x \Big[ -{1\over 2}\epsilon(x^--y^-) 
		\delta^2(x_\bot -y_\bot) V_\alpha^-(x|y) \Big] \nonumber \\
		&&  + \partial^i_x \Big[{1\over 2}\epsilon(x^--y^-)
		\delta^2(x_\bot -y_\bot)\Big[V_\alpha^i(x|y) \nonumber \\
		&& ~~~~~~~~~~~~~~~~~ ~~~ + i 
		\epsilon^{ij}A_\alpha^j(x|y)\Big]\Big] - h.c. \Bigg\} \, ,
		\label{c+-c}
\end{eqnarray}
where $V_\alpha^\mu$ and $A_\alpha^\mu$ are defined as the bilocal 
vector and axial vector currents: 
\begin{eqnarray}
	V_\alpha^\mu(x|y) &=& \overline{\psi}_\alpha(x) \gamma^\mu 
		\psi_\alpha(y) \, , \\
	A_\alpha^\mu(x|y) &=& \overline{\psi}_\alpha(x) \gamma^\mu 
		\gamma_5 \psi_\alpha(y) \, . 
\end{eqnarray}

As we can see, the light-front current commutators are very different 
from the equal-time current commutators. Here the commutator is indeed 
given by terms involving spatial derivatives. These  space-derivatives
come from the non-locality of $\psi_-(x)$ on the light-front. In the 
equal-time formulation, there is no such nonlocality to the fermion 
field.  Therefore one cannot derive such a commutator from the naive 
canonical equal-time commutations. As we will soon see in the next
section it is these spatial derivatives that lead to the simple 
expressions of the structure functions in terms of bilocal current 
matrix elements.   This is  an 
essential feature in the present approach that make the light-front 
current algebra specially useful in the exploration of the deep 
inelastic structure functions.

The commutators for other current components can also be found 
straightforwardly. For example, 
\begin{eqnarray}  
	\Big[ J^+(x)~, &~& J^i(y) \Big]_{x^+=y^+} = \sum_\alpha e_\alpha^2
		\Bigg\{ \partial^+_x \Big[ -{1\over 2}\epsilon(x^--y^-) 
		\delta^2(x_\bot -y_\bot) V_\alpha^i(x|y) \Big] \nonumber \\
		& & + \partial^j_x \Big[{1\over 2}\epsilon(x^--y^-)
		\delta^2(x_\bot -y_\bot)\Big(g^{ij}V_\alpha^+(x|y) + i 
		\epsilon^{ij}A_\alpha^+(x|y)\Big)\Big] - h.c. \Bigg\} \, .
		\label{c+ic}
\end{eqnarray}
Thus, one can use Eq.(\ref{c+-c}) to extract the structure
functions and then use Eq.(\ref{c+ic}) to make a consistency check.
	
Now, the Compton scattering amplitude in the large $q^-$ limit can be
immediately expressed in terms of the hadronic matrix elements of the
bilocal vector and axial vector currents. For example, the $(+-)$
component $T^{+-}$ is given by 
\begin{eqnarray}
	T^{+-} & \stackrel{{\rm large}~q^-}{=} & -{1\over q^+} 
		\int d\xi^- e^{iq^+\xi^-/2} \epsilon
		(\xi) \langle PS| \sum_\alpha e_\alpha^2 \Big\{ {i\over 2} 
		q^+ V_\alpha^-(\xi^- |0)  \nonumber \\
	& & ~~~~~~~~~~~~~ - {i \over 2} q_\bot^i [ V_\alpha^i (\xi^-|0) 
		+ i \epsilon^{ij} A_\alpha^j(\xi^-|0)] \Big\}- h.c.
		~ | PS \rangle  \, . \label{t+-}
\end{eqnarray} 
The above result shows that the bilocal vector and axial vector 
currents entering in $T^{\mu \nu}$ are separated only in the 
longitudinal direction $\xi^- = \xi^0 - \xi^3$. This property 
naturally leads to the well-known scaling behavior of the 
structure functions 
when we ignore the QCD dynamics at very high $Q^2$. In the next 
section, we will extract explicitly the structure functions in 
terms of these bilocal current matrix elements.
%%%%%%%%%%%%%%%%%%%%%%%%%%%%%%%%%%%%%%%%%%%%%%%%%%%%%%%%%%%%%
\section{The generalized expressions for Deep inelastic structure functions}
%%%%%%%%%%%%%%%%%%%%%%%%%%%%%%%%%%%%%%%%%%%%%%%%%%%%%%%%%%%%%
Now, let us pick up the same $(+-)$ component of the hadronic
tensor Eq.(\ref{wfun}) to find the deep inelastic structure
function by comparing with Eq.(\ref{t+-}) through Eq.(\ref{twr}).
In the large $q^-$ limit,
\begin{eqnarray}
	W^{+-} &=& {1\over 2}F_L +  (P_\perp)^2 {F_2 \over \nu}
                  - 2 P_\bot \cdot q_\bot{F_2\over q^2} \nonumber \\
	       & & ~~~~~~~~~ + 2i\epsilon^{ij}q_i
		\Big[S_{jL} {g_1\over \nu} + S_{jT}{g_T\over \nu} \Big] \, , 
		\label{w+-lq}
\end{eqnarray}
where $S_{jT} = S_j - S^+ {P_j \over P^+}$ and $S_{jL} = S_j -S_{jT}
= S^+ {P_j \over P^+}$, and  $\nu = {1\over 2}P^+q^-$ in the large $q^-$ 
limit. We introduce the form factors for the 
bilocal current matrix elements,
\begin{eqnarray}  
	\langle PS| V_\alpha^\mu(\xi|0) - V^\mu_\alpha(0|\xi) |PS \rangle 
		&=& P^\mu \overline{V}_{1\alpha}(P^2,\xi \cdot P) +
		\xi^\mu \overline{V}_{2 \alpha}(P^2,\xi \cdot P) \, , 
		\label{bcff} \\
	\langle PS| A_\alpha^\mu(\xi|0) + A^\mu_\alpha(0|\xi) |PS \rangle 
		&=& S^\mu \overline{A}_{1\alpha}(P^2,\xi \cdot P) +
		P^\mu \xi \cdot S \overline{A}_{2\alpha}(P^2,\xi \cdot P) 
		\nonumber \\
	& & ~~~~~~~~~~~~~ + \xi^\mu S \cdot \xi \overline{A}_{3\alpha}(P^2,
		\xi \cdot P) \, . \label{bacff}
\end{eqnarray}
Since, $\xi^{+, \perp} =0$,  it follows that the matrix elements of the plus
and transverse components of the bilocal current yield the same form factor
$\overline{V}_{1\alpha}$. 
Using the definition
\begin{equation}
	\epsilon(\xi^-) = - {i \over \pi} \int_{-\infty}^\infty
		{d q^+ \over q^+} e^{i q^+\xi^-/2} \, ,
\end{equation} 
we find that
\begin{eqnarray}
	T^{+-} = -{1 \over \pi q^-} \int_{-\infty}^\infty {d{q'}^+ \over {q'}^+ 
		- q^+ } \int_{-\infty}^\infty & & d\xi^- e^{iq^+\xi^-/2}
		\sum_\alpha e_\alpha^2 \Bigg\{{1\over 2}(P^-q^+ - P_\bot
		\cdot q_\bot) \overline{V}_{1 \alpha} \nonumber \\
	& & + {1\over 2} q^+ \xi^- \overline{V}_{2 \alpha} + {i\over 2}
		\epsilon^{ij}q_i \Big[S_j \overline{A}_{1 \alpha} + P_j
		{S^+ \xi^-\over 2} \overline{A}_{2 \alpha} \Big] \Bigg\} 
		\, . \label{t+-lq}
\end{eqnarray}
The bilocal current form factors are determined from Eq.(\ref{bcff}):
\begin{eqnarray}
	\overline{V}_{1\alpha} &=& {1\over P^+} \langle PS |
		\overline{\psi}_\alpha (\xi^-) \gamma^+ \psi_\alpha (0) - 	
		\overline{\psi}_\alpha (0)\gamma^+ \psi_\alpha (\xi^-) |PS 
		\rangle \label{v1+} \\
	&=& {1\over P^i} \langle PS | \overline{\psi}_\alpha (\xi^-) 
		\gamma^i \psi_\alpha (0) - \overline{\psi}_\alpha (0)
		\gamma^i \psi_\alpha(\xi^-) |PS \rangle \, , \label{v1i} \\
	\overline{V}_{2\alpha} &=& {1\over \xi^-} \langle PS | 
	      \overline{\psi}_\alpha (\xi^-) \Big(\gamma^--{P^-\over P^+}
		\gamma^+ \Big)\psi_\alpha (0) - h.c. |PS \rangle \nonumber \\
	&=& {1\over \xi^-} \langle PS | \overline{\psi}_\alpha (\xi^-) 
		\Big(\gamma^--{P^-\over P^i}\gamma^i \Big)
		\psi_\alpha (0) - h.c. |PS \rangle \, , \\
	\overline{A}_{1\alpha} &=& {1\over S^i_T} \langle PS| 
		\overline{\psi}_\alpha (\xi^-) \Big(\gamma^i
		-{P^i\over P^+}\gamma^+ \Big)\gamma_5\psi_\alpha (0) 
		+ h.c. |PS \rangle \, , \\
	\overline{A}_{2\alpha} &=& {-2\over P^+\xi^- S^i_T} \langle PS| 
		\overline{\psi}_\alpha(\xi^-) 
		\Big(\gamma^i -{S^i\over S^+}\gamma^+ \Big)
		\gamma_5 \psi_\alpha(0) + h.c. |PS \rangle \, . \label{a2}
\end{eqnarray}
Comparing with Eqs.(\ref{w+-lq}) and (\ref{t+-lq}) through Eq.(\ref{twr}), 
we obtain, with $\eta \equiv {1\over 2}P^+ \xi^-$, 
\begin{eqnarray}
	{F_2(x,Q^2)\over x} &=& {1\over 4\pi} \int d\eta e^{-i\eta x} 
		\sum_\alpha e^2_\alpha \overline{V}_{1\alpha} 
		\label{f20} \\
	&=& {1\over 4\pi P^+} \int d\eta e^{-i\eta x} \sum_\alpha 
		e^2_\alpha \langle PS| \overline{\psi}_\alpha (\xi^-) 
		\gamma^+ \psi_\alpha (0) - \overline{\psi}_\alpha (0)
		\gamma^+ \psi_\alpha (\xi^-) |PS \rangle  \label{f2+} \\
	&=& {1\over 4\pi P^i_\bot} \int d\eta e^{-i\eta x} \sum_\alpha
                e^2_\alpha \langle PS | \overline{\psi}_\alpha (\xi^-) 
		\gamma_\bot^i \psi_\alpha (0) - \overline{\psi}_\alpha (0)
		\gamma_\bot^i \psi_\alpha(\xi^-) |PS \rangle \, , \label{f2i}
\end{eqnarray}
where the last equality is found for the first time here \cite{Hari97}. 
Its physical interpretation will be given later.
\begin{eqnarray}
	F_L(x,Q^2) &=& -{q^+\over \pi P^+q^-} \int d\eta e^{-i\eta x} 
		\sum_\alpha e^2_\alpha \Big[(P^- - {P_\perp^2 \over P^+}) 
		\overline{V}_{1\alpha} + \xi^- \overline{V}_{2\alpha} \Big ]
		\nonumber \\
	&=& {P^+\over 4\pi } \Bigg({2x\over Q}\Bigg)^2 \int d\eta 
		e^{-i\eta x} \sum_\alpha e^2_\alpha \langle PS| 
		\overline{\psi}_\alpha (\xi^-) \nonumber \\
	& & ~~~~~~~~~~~~~~~~~~~~~~~~~  \times 
		\Big(\gamma^--{P_\perp^2 \over (P^+)^2}
                       \gamma^+ \Big)\psi_\alpha 
		(0) - h.c. |PS \rangle  \label{fl}
%	&=& {1\over 4\pi P^+} \Bigg({2x\over Q}\Bigg)^2 
%		\int d\eta e^{-i\eta x} \sum_\alpha 
%		e^2_\alpha \langle PS| \overline{\psi}_\alpha (\xi^-) 
%		\Big(\gamma^--{P^-\over P_\bot^i}\gamma^i_\bot \Big)
%		\psi_\alpha (0) - h.c. |PS \rangle \, , 
\end{eqnarray}
where the first equality may be reduced to the same expression obtained 
by the collinear expansion in the Feynman diagrammatic method up to the 
order twist-four \cite{EPF83}. But here it is directly obtained in the 
leading order in the $1/q^-$ expansion without involving the concept 
of twist expansion.  
Moreover, the polarized structure functions can also be found directly as 
\begin{eqnarray}
	g_1(x,Q^2) &=& {1\over 8\pi} \int d\eta e^{-i\eta x} \sum_\alpha
		e^2_\alpha \Big(\overline{A}_{1\alpha} + {1 \over 2}
           P^+ \xi^- \overline{A}_{2\alpha} \Big) \label{g10} \\
	&=& {1\over 8 \pi S^+} \int d\eta e^{-i\eta x} \sum_\alpha
		e^2_\alpha \langle P S| \overline{\psi}_\alpha (\xi^-)
		\gamma^+ \gamma_5 \psi_\alpha(0) + \overline{\psi}(0) 
		\gamma^+\gamma_5 \psi(\xi^-) |PS \rangle , \label{g1}\\
	g_T(x,Q^2) &=& {1 \over 8\pi} \int d\eta e^{-i\eta x} \sum_\alpha
		e^2_\alpha \overline{A}_{1\alpha} 
		\label{gt0}\\
	&=& {1\over 8\pi S^i_T} \int d\eta 
		e^{-i\eta x} \sum_\alpha e^2_\alpha \langle PS|\overline{
		\psi}_\alpha(\xi^-) \Big(\gamma^i -{P^i\over P^+}
		\gamma^+ \Big)\gamma_5 \psi_\alpha(0) +~ h.c. |PS \rangle 
		\, . \label{gt}
\end{eqnarray}

The above results are derived without recourse to perturbation theory,
and also without the use of concept of collinear and massless partons.
They are also the most general expressions for the leading contribution
(in the $1/q^-$ expansion, not the leading contribution in terms of 
twists) to the deep inelastic structure functions in which the target
is in an arbitrary Lorentz frame. Some of these expressions have not 
ever been obtained in previous works. The $x$-dependence of these 
structure functions is obvious in the above expressions. The scale 
($Q^2$)-dependence is 
hidden in the hadronic bound states $|PS \rangle$ which can be 
described by multi-parton wave functions. In the next section,
we shall analyze the complexities of structure
functions in terms of bound states in our description.

%%%%%%%%%%%%%%%%%%%%%%%%%%%%%%%%%%%%%%%%%%%%%%%%
\section{Complexities of structure functions}
%%%%%%%%%%%%%%%%%%%%%%%%%%%%%%%%%%%%%%%%%%%%%%%%
\subsection{Multi-parton wave functions}

As we have seen, the derivation of structure functions in the previous 
section is apparently quite different from the QCD improved (or 
field theoretical) parton
model. The latter which is based on the collinear concept is purely 
an application of perturbation theory where further exploration 
of nonperturbative dynamics is lacking. Our description is also 
obviously very 
different from the OPE method. On the other hand, structure functions 
themselves are dominated by the nonperturbative quark-gluon dynamics.
When we formulate them on the light-front, the structure functions
are proportional to the simple hadronic matrix elements of the bilocal
currents that are separated only in the longitudinal direction. 
In this formulation, no time evolution or propagation is explicitly 
involved in the matrix elements. Hence, unlike the OPE or the
perturbative field theory descriptions of parton model, all the 
perturbative and nonperturbative  dynamics here are completely carried 
by the structure of target's bound state. This is closer to the real 
physical picture probed in experiments. 

The bound state of a hadron on light-front can be simply 
expanded in terms of the Fock states,
\begin{equation}
        |PS \rangle = \sum_{n,\lambda_i} \int' dx_i d^2\kappa_{\bot i} 
		 | n, x_iP^+,x_iP_{\bot}+ 
		\kappa_{\bot i}, \lambda_i \rangle \Phi^S_n 
		(x_i,\kappa_{\bot i}, \lambda_i) \, , \label{lfwf} 
\end{equation}
where $n$ represents $n$ constituents contained in the Fock state 
$|n, x_i P^+, x_i P_{\bot} + \kappa_{\bot i}, \lambda_i \rangle$, 
$\lambda_i$ is the helicity of the i-th constituent, $\int'$ denotes 
the integral over the space:
\begin{equation}  \label{lfspc}
        \sum_i x_i = 1, ~~ {\rm and} ~~~ \sum_i \kappa_{\bot i} = 0
\end{equation}
while $x_i$ is the fraction of the total longitudinal momentum 
carried by the $i$-th constituent, and $\kappa_{\bot i}$ is its relative 
transverse   momentum with respect to the center mass frame:
\begin{equation}
        x_i = { p_i^+ \over P^+}~~, ~~~ \kappa_{i\bot} = p_{i\bot} - x_i 
P_{\bot} \end{equation}
with $p_i^+, p_{i\bot}$ the longitudinal and transverse momenta
of the $i$-th constituent. $\Phi^S_n (x_i,\kappa_{\bot i},\lambda_i)$ 
is the amplitude of the Fock state $| n, x_iP^+,x_iP_{\bot}+ \kappa_{\bot 
i},\lambda_i \rangle $, i.e., the {\it multi-parton wave function},
which is boost invariant and satisfy the normalization condition: 
\begin{equation}
        \sum_{n,\lambda_i} \int' dx_i d^2\kappa_{\bot i} 
		|\Phi^S_n (x_i,\kappa_{\bot i},\lambda_i)|^2 = 1,
\end{equation}
and is, in principle, determined from the light-front bound state 
equation,
\begin{equation}
        \Big(M^2 - \sum_{i=1}^n { \kappa_{i\bot}^2 + m_i^2 \over x_i} 
		\Big) \left[\begin{array}{c} \Phi^S_{qqq} \\
                \Phi^S_{qqqg} \\ \vdots \end{array} \right]
                  = \left[ \begin{array}{ccc} \langle qqq
                | H_{int} | qqq \rangle & \langle qqq | H_{int}
                | qqqg \rangle & \cdots \\ \langle qqq g
                | H_{int} | qqq \rangle & \cdots & ~~  \\ \vdots &
                \ddots & ~~ \end{array} \right] \left[\begin{array}{c}
                \Phi^S_{qqq} \\ \Phi^S_{qqqg} \\ \vdots \end{array}
                \right] . \label{lfbe}
\end{equation}
Here $H_{int}$ is the interaction part of the light-front QCD 
Hamiltonian \cite{zhang93}.
Thus, the complexities of the structure functions carried by 
hadronic bound states are now translated into the language of 
multi-parton wave functions on the light-front, rather than 
composite operators in OPE.

Explicitly, let us look at the structure function $F_2(x,Q^2)$. It is 
found for the first time by us \cite{Hari97} that $F_2$ can be expressed
in terms of  
a matrix element of either the plus component or the transverse 
component of the bilocal vector current. On the light-front, these 
two components have totally different operator structures but amazingly
their matrix elements determine the same structure function. 

The plus component (usually called the ``good" component),
\begin{equation}
	\overline{\psi} \gamma^+ \psi = 2 \psi_+^\dagger \psi_+ \, ,
\end{equation}	
has no explicit dynamical dependence, and has the lowest mass 
dimension (a twist-two operator in OPE language). The corresponding 
matrix element has straightforward parton interpretation. It is
clear from the above operator that on the light-front it is just
a quark (parton) number operator which immediately leads to the
fact that $F_2$ is proportional to parton density distributions
$q_\alpha (x,Q^2)$. 
\begin{eqnarray}
	{F_2(x,Q^2) \over x} &=& \sum_\alpha e^2_\alpha q_\alpha
		(x,Q^2) \, , \label{f2part} \\
	q_\alpha (x,Q^2) &=&  \int d^2 k_\bot 
		\langle PS | \sum_\lambda b_\alpha ^\dagger(k,\lambda) 
		b_\alpha (k,\lambda) | PS \rangle  \nonumber \\
	&=& \int d^2 \kappa_\bot \sum_{n,\lambda_i} \int'' dx_i d^2 
		\kappa_{\bot i} |\Phi^S_{n, \alpha} (x, x_i, 
		\kappa_{\bot i},\lambda_i)|^2 \, , \label{partd} 
\end{eqnarray}
where the $Q^2$-dependence is 
carried by the multi-parton wave functions with the active parton 
renormalized at the scale $Q^2$, $\int''$ 
denotes the integral in the right-hand-side over the space of Eq.
(\ref{lfspc}) except for the active parton $(x, \kappa_\bot)=
(k^+/P^+, p_\bot - xP_\bot)$. With this consideration it is 
straightforward to derive the logarithmic corrections that is the 
same as that obtained in the QCD improved parton model or in the 
OPE, as will be given in \cite{deep2}. In this case, all the three 
descriptions are almost the same. 
The only difference here is that in our framework, 
the perturbative QCD dynamics is transferred from the composite 
operator into the scale-dependent multi-parton wave functions 
on the light-front, which enables us to describe the nonperturbative 
dynamics in the same framework.
 
But the transverse component (sometimes called the ``bad" component), 
\begin{equation}
	\overline{\psi} \gamma^i \psi = \overline{\psi}_- \gamma^i_\bot 
		\psi_+ + \overline{\psi}_+ \gamma^i_\bot \psi_- \, ,
\end{equation}	
depends explicitly on the fundamental quark-gluon interaction 
in QCD. According to the twist analysis \cite{jaffeji}, 
the transverse component of the bilocal current is a twist-three
operator which has no simple parton interpretation. However, we have
explicitly shown \cite{Hari97} that the corresponding matrix 
element of the above transverse component must have the same parton 
interpretation as that from the plus component. This is indeed
obvious because they represent the same form factor of the 
bilocal current [see Eqs.(\ref{v1+}) and (\ref{v1i})] and they 
describe the same structure function $F_2$ [see Eqs.(\ref{f2+})
and (\ref{f2i})]. This conclusion is very different from the 
current understanding of the twist of composite operators 
in the OPE or in the QCD improved parton model.

The explicit calculations in Ref.\cite{Hari97} further demonstrate 
that the real dynamics contained in the structure functions 
is determined by the matrix element with the rich information
carried by the multi-parton wave functions. It is the complicated 
multi-parton wave functions that causes the same behavior for
the matrix elements of 
the two apparently different operators (the plus and transverse 
components of bilocal current). However, this 
property was overlooked in previous works. Partly because one 
usually worked in a specific Lorentz frame, such as the rest 
frame or the infinite momentum frame in which the target's 
transverse momentum $P_\bot = 0$, and also because 
enough attention was not paid on the hadronic bound state structure. 
With  $P_\bot=0$, $F_2$ can only be expressed by Eq. (\ref{f2+}). 
The expression in terms of the matrix element of the transverse 
component of the bilocal current, i.e. Eq. (\ref{f2i}), does 
not exist if one lets $P_\bot=0$. In other words, picking up a
specific Lorentz frame on light-front may lead to some ambiguities.
In the next subsection, we shall present some discussion
on such ambiguity in the concept of twist which is currently
one of the most interesting topics in the study of deep inelastic
processes.

%%%%%%%%%%%%%%%%%%%%%%%%%%%%%%%%%%%%%%%%%%%%%%%%%%%%%%%%%%%%
\subsection{A Lorentz invariant definition of the twist}
%%%%%%%%%%%%%%%%%%%%%%%%%%%%%%%%%%%%%%%%%%%%%%%%%%%%%%%%%%%%
As we have seen the lack of the additional expression (\ref{f2i})
for $F_2$ in other approaches may be due to the 
specific choice of the Lorentz frame one often used.
Indeed, since the deep inelastic processes are dominated by the 
physics close to the light-front, one often choose the following
parameterization to analyze the structure functions
\cite{pqcd,jaffeji}:
\begin{eqnarray}
	P^\mu &=& p^\mu + n^\mu \, , \nonumber \\
	q^\mu &=& {1\over M^2}\Big(\nu - \sqrt{\nu^2 + M^2Q^2} 
		\Big) p^\mu + {1\over 2}\Big( \nu + \sqrt{\nu^2 
		+ M^2Q^2} \Big) n^\nu	\, ,  \label{lcc}
\end{eqnarray}
where $p^\mu = (p, 0, 0, p)$ and $n^\mu=(2, 0, 0, -2)$ are the 
light-like vectors;  $p=M/2$ yields the target rest frame and
$p \rightarrow \infty$ leads to the infinite momentum frame. 
This choice of Lorentz frame for the deep inelastic processes
may simplify the analysis, but {\it it loses the generality}
when one tries to extract some general conclusions. A typical
example is the twist analysis in the determination 
of the dominant contributions to deep inelastic structure functions. 

The dominant contribution to the structure functions in terms of 
twist expansion is the basis of the OPE and QCD improved 
parton model analysis. Measuring the higher-twist contributions has 
become a main topic in the current and future experiments since
it may provide some nontrivial test of perturbative QCD beyond
the logarithmic corrections. However, the definition of twist in 
literature is not unique and its physical interpretation is also 
not very transparent. According to standard definition,
twist of a given composite operator is given by mass dimension of 
the operator minus its spin. In reality, this definition  
has no direct connection with experimental parameters. 

Practically, the twist can be given in a less formal way:
the twist of an invariant matrix element of a light-front bilocal 
operator, which controls the scaling behavior of the matrix 
element is determined by the power of $1/Q$ that it contributes to
deep inelastic processes \cite{jaffeji}. This definition is
correct if the {\it invariant matrix elements} here are specified
by those Lorentz invariant form factors of the bilocal currents 
given in Eqs.(\ref{v1+}-\ref{a2}). However, in previous applications, 
one used the specific coordinates 
(\ref{lcc}) to define the invariant matrix elements of the 
light-front bilocal currents and further used  the usual 
dimensional analysis to determine the corresponding twists. 
Unfortunately, the twist of an matrix element defined 
in such specific coordinates with the usual dimensional analysis 
is not Lorentz invariant and contains ambiguous. 

For example, according to the above definitions, Jaffe and Ji 
concluded that plus component of bilocal current is a 
twist-two operator, and transverse and minus components 
are twist three and four operators \cite{jaffeji}.
%These results appear consistent with the twist analysis obtained
%in QCD improved parton model \cite{EPF83}. In QCD 
%improved parton model, one uses the same coordinates of (\ref{lcc})
%and then simply counts the number of individual 
%light-front quark and gluon field lines linking the hard and 
%soft parts as the degree of twist for Feynman diagrams 
%in contributing to hard processes. 
%However, the problem is that the twist thus defined 
%is not Lorentz invariant and the suppression by $1/Q$ does not 
%always work in hard processes. This can be seen from Eq.(\ref{f2i}).
%Based on the conventional definition of twist, 
As a result, matrix elements of the transverse component of  
bilocal currents should be suppressed by a power of $1/Q$ in 
comparison to the plus component matrix element. 
However, we find that the transverse component matrix element is 
actually suppressed by $P_\bot$ rather than $Q^2$: Smaller the 
value $P_\bot$ takes in a specific  choice of a Lorentz frame for 
the target, smaller will be the contribution arising from the   
transverse component matrix element. In the infinite 
momentum frame or in the target rest frame where the target has 
zero $P_\bot$, the transverse component matrix element must vanish. 
On the other hand, the plus component matrix element retains its 
value, no matter what the $Q^2$ value is (modulo logarithms).  
Thus, one cannot obtain the conclusion that transverse component 
of bilocal current is a twist-three operator that is suppressed by 
the factor $1/Q$ in hard processes. The fact that the transverse
component matrix element vanishes is a consequence of using the 
specific Lorentz frames with $P_\bot =0$ and is independent of the 
value of $Q^2$.

Indeed, we also find that twist of Lorentz invariant matrix elements 
of light-front bilocal currents can be properly defined {\it in terms 
of light-front power counting analysis} for light-cone
dominated deep inelastic processes.  Here one needs to consider the 
target in an  arbitrary Lorentz frame. Then Eqs.(\ref{bcff})-(\ref{bacff}) 
give the general definition of the Lorentz invariant matrix elements, 
i.e., the form factors of the bilocal vector and axial vector currents. 
The twist of a Lorentz invariant matrix element of bilocal currents is 
thus directly given by the mass dimension of the corresponding bilocal
current form factors in light-front power counting.
One must notice that {\it light-front power counting} is very different
from the usual one \cite{Wilson94}. The light-front
power counting separately introduces different scales for  
transverse and longitudinal dimensions. Only the transverse dimension 
has mass scale,
\begin{equation}
	m ~~ : ~~~ {1\over x_\bot} \, .
\end{equation}
In light-front power counting \cite{Wilson94}, 
both the light-front quark field $\psi_+$ and the gluon field 
$A_\bot$ have mass dimension one because their power assignments
are given by
\begin{equation}
	\psi_+ ~~: ~~~{1\over \sqrt{x^-}} {1 \over x_\bot}~, ~~~~
	~A_\bot ~~: ~~~{1 \over x_\bot}  \, ,
\end{equation}
although their longitudinal dimensions are different. Besides, the
transverse derivative has mass dimension one and the longitudinal 
derivative has no mass dimension. Thus, from Eq. (\ref{qkmc}), we
can see that $\psi_- ~:~ {\sqrt{x^-} \over x_\bot^2}$ has mass dimension $2$. 

In deep inelastic scattering processes, we have shown in Eqs.(\ref{bcff})
and (\ref{bacff}) that there are five Lorentz invariant form factors 
associated with the matrix elements of the bilocal vector and 
axial vector currents: $\overline{V}_{1\alpha}$, $\overline{V}_{2\alpha}$,
 $\overline{A}_{1\alpha}$,  $\overline{A}_{2\alpha}$ and  
$\overline{A}_{3\alpha}$. With simple power counting, one can
easily find that three of them, ${V}_{1\alpha}$, 
 $\overline{A}_{1\alpha}$ and  $\overline{A}_{2\alpha}$, are 
twist-two, while the remaining two,
$\overline{A}_{2\alpha}$ and $\overline{A}_{3\alpha}$, are
twist-four. Based on the normalization of bound state
\begin{equation}
	\langle P'S' | PS \rangle = 2(2\pi)^3 P^+ \delta^3(P-P')
		\delta_{SS'} \, ,
\end{equation}
the light-front mass dimension of the state $| PS \rangle $ is $-1$.
Then we can easily count the mass dimension for each bilocal current
form factor in light-front power counting as follows: 
\begin{center}
\begin{tabular}{c c c c c c c c}
	$\overline{V}_1$ & ~$\sim$~ & $\langle PS |$ & $\psi^\dagger_+$
		&~$\psi_+$ & $| PS \rangle$ & &\\
	m-dim($\overline{V}_1$) & $=$ & $-1$ & $+1$ & $+1$ & $-1$ & $= 0$   
		\, , & \\
	~~~~~ or ~ & & & & & & & \\ 
	   &$\sim$ & ${1\over P_\bot}$ & $\langle PS |$ & $\psi^\dagger_-$&
		$\psi_+$ & $|PS \rangle$ &  \\
 		& $=$ & $-1$ & $-1$ & $+2$ & $+1$ & $-1$& $= 0$ \, , \\
	& & & & & & & \\
       $\overline{A}_1$ &~ $\sim$~ & ${1\over S_T}$ & $\langle PS|$ &
                $~\psi^\dagger_-$ & $\psi_+$ & $|PS \rangle$ &  \\
	m-dim($\overline{A}_1$) &$=$& $-1$ & $-1$ & $+2$ & $+1$ & $-1$& $=0$ 
		\, , \\
	& & & & & & & \\
        $\overline{A}_2$ &~ $\sim$~ & ${1\over S_T}$ & $\langle PS|$ &
                $~\psi^\dagger_-$ & $\psi_+$ & $|PS \rangle$  \\
	m-dim($\overline{A}_2$) &$=$& $-1$ & $-1$ & $+2$ & $+1$ & $-1$ & $=0$ 
		\, ,  
\end{tabular}
\end{center}
\begin{center}
\begin{tabular}{c c c c c c c}
        $\overline{V}_2$ &~ $\sim$~ & $\langle PS |$ & $\psi^\dagger_-$ & 
		$~\psi_-$ & $|PS \rangle$ &  \\
	m-dim($\overline{V}_2$) & $=$ & $-1$ & $+2$ & $+2$ & $-1$ & $=2$ \, ,\\
	& & & & & & \\
        $\overline{A}_3$ &~ $\sim$ ~& $\langle PS|$ & $\psi^\dagger_-$ & 
		$\psi_-$ & $|PS \rangle$ & \\
	m-dim($\overline{A}_3$) &$=$& $-1$ & $+2$ & $+2$ & $-1$ & $=2$ \, .
\end{tabular}
\end{center}
The twist of the Lorentz invariant matrix elements (i.e. the bilocal
current form factors) is defined as its mass dimension plus 2. 
Therefore, $\overline{V}_{1\alpha}$, $\overline{A}_{1\alpha}$ 
and  $\overline{A}_{2\alpha}$ have twist-two, and 
$\overline{V}_{2\alpha}$ and $\overline{A}_{3\alpha}$ are twist-four
Lorentz invariant matrix elements.

As a matter of fact, there is indeed no twist-three Lorentz invariant 
matrix element in the leading contributions to deep-inelastic 
lepton-hadron processes. This conclusion should not be very surprising 
since experimentally no twist-three matrix element with a ${1/Q}$
coefficient has been extracted in deep inelastic processes.
%in the sense 
%that no evidence has indicated to have the $M/Q$ suppression 
%in experimental data. 
Theoretically, it is also hard to find 
some result with power correction proportional to $1/Q$ rather 
than $1/Q^2$ in inclusive lepton-hadron deep inelastic processes. 
Our analysis shows that the next higher twist 
corrections to the leading twist-two contributions in deep 
inelastic processes are twist-four contributions which 
should be suppressed by the factor $1/Q^2$. Indeed, from 
light-front power counting analysis, the mass dimension of 
Lorentz invariant bilocal current form factors is much more meaningful 
than the conventional definition of twist.  If the light-front mass 
dimension of a form factor is zero, it means that there is no $1/Q$ 
power suppression, see Eqs.(\ref{f20}), (\ref{g10}), and \ref{gt0})
for $F_2, g_1$ and $g_T$. 
If the mass dimension of a form factor is not zero, then there might
be some ${\cal O}(1/Q)$ power suppression to ensure that the structure 
function has the same dimensions [cf. Eq.(\ref{fl}) for $F_L$]. 
%It seems that the concept of twist is not necessary useful.

Now we can see that because the Lorentz invariant form factor
$\overline{V}_{1\alpha}$ is a twist-two matrix element,
no matter if it is expressed in terms of the matrix element 
of plus or transverse component of the bilocal vector current, 
the leading contribution 
of $F_2$ is always a twist-two contribution, as is 
already known. The contribution of the transverse component
matrix element goes to zero when we analyze it in the specific light-cone
coordinates (\ref{lcc}), but the ratio of the transverse component
of the matrix element and $P_\bot$, which is proportional to $F_2$,
remains invariant, as a result of  Lorentz invariance. Thus the possible 
inconsistency of eqs.(\ref{f2+}) and (\ref{f2i}) in terms of the
naive definition of twist does not really exist. It appears
that the conventional definition of twist that is currently used
in the literature may not be proper in the analysis of deep inelastic 
processes. 

It is also to be noted that the twist analysis is meaningful only 
in some restricted portion of the phase space. For example, in the 
region of phase space connected to the limit $x \to 1$, that is when 
one of the particle in the Fock state is carrying almost all the 
hadron momenta, the states are far off its energy shell. Apparently, 
it seems that the contribution from such states to the structure
functions are power suppressed, for it needs multiparton interaction 
in the first place to dump all the longitudinal momenta to one 
parton to produce such a Fock state. Yet their contributions are 
comparable to the leading twist ones where center of mass energy 
remains fixed. In fact, these contributions are more important and 
controls the behaviors of structure functions in the $x \to 1$ 
limit \cite{Brodsky1,Brodsky2}. In any case, as shown in these 
references, the twist expansion looses its significance all together 
in this limit.

Next, we shall analyze the twist of polarized structure functions, 
which has remained ambiguous for many years in the literature.

\subsection{Twist analysis of the polarized structure functions}

With the above analysis of the complexities of multi-parton
wave functions and the new understanding of the concept of 
twist, we now explore the properties of the polarized structure 
functions $g_1$ and $g_T$.

In the literature, the properties of $g_2$ are usually discussed rather 
than the real transverse polarized structure function $g_T$ for 
historical reasons. But as we have pointed out, $g_2$ by itself has no 
real physical meaning. However, to clarify the confusion about the
contributions of different twists to $g_2$ in the literature, we focus our
attention on $g_2$ here. The polarized structure functions, especially 
the transverse polarized structure function, have recently received 
much theoretical and experimental attention. The polarized structure 
functions $g_1$ and $g_2$ are of great interest due to the fact that the 
former has been considered the best quantity to understand the 
origin of the proton spin, while the later may lead to the 
first observation of higher twist (twist-three) contribution 
in DIS. In the early naive parton model, Feynman claimed that 
$g_2$, just like $g_1$, has a simple parton interpretation because 
$g_T(x) = g_1(x) + g_2(x) = {1\over 2} 
\sum_q e_q^2 \Delta q_T(x)$, where $\Delta q_T(x)$ is the distribution of
transversely polarized quarks in a transversely polarized 
nucleon \cite{Feynman72}. Wandzura and Wilczek 
studied the properties of $g_2$ in OPE analysis \cite{Wandzura77}
and they claimed that except for a twist-three contribution 
$\overline{g}_2(x)$ which may be negligible in their model 
calculation, $g_2(x)$ can be related to an integral over $g_1(x)$:
\begin{equation} \label {g223}
        g_2(x) = g_2^{WW}(x) + \overline{g}_2(x), ~~~~~~~~
        g_2^{WW}(x) = - g_1(x) + \int_x^1 {dy\over y} g_1(y) .
\end{equation}
The relation between $g_2^{WW}$ and $g_1(x)$ is called
Wandzura-Wilczek (WW) relation. Later, Shuryak and 
Vainshtein pointed out that the twist-three contribution 
$\overline{g}_2$ is a direct quark-gluon interaction
effect which is important for $g_2(x,Q^2)$ \cite{Chuyk84} so that
the measurement of $g_2$ may also be very sensitive to the
interaction dependent higher twist effects in QCD \cite{jaffeji}.

Here we shall apply light-front power counting to the analysis
of the twist of  
polarized structure functions. The operator involved in $g_1$ is 
relatively simple, it is the plus component of bilocal axial 
vector current,
\begin{equation}
	\overline{\psi}_\alpha (\xi^-) \gamma^+ \gamma_5 \psi_\alpha
		= 2 \psi_+^\dagger(\xi^-) \gamma_5 \psi_+(0) \, ,
\end{equation}
which is a twist-two operator in the conventional definition. The
corresponding Lorentz invariant matrix element is also twist-two
in our definition.  However, the 
operator determining the transverse polarized structure function 
$g_2$ is much more complicated. From Eqs. (\ref{g1}) and (\ref{gt}), 
one can obtain: 
\begin{eqnarray}
	g_2(x,Q^2) &=& g_T(x,Q^2) - g_1(x,Q^2) = - {1 \over 8\pi}
		\int d\eta e^{-i\eta x} \sum_{\alpha} e_\alpha^2 
		{P^+ \xi^- \over 2} \overline{A}_{2\alpha} \nonumber \\  
  	&=& {1\over 8\pi S^i_T} \int d\eta e^{-i\eta x} \sum_\alpha 
		e^2_\alpha \langle PS| \Big[ \overline{\psi}_\alpha(\xi^-) 
		\Big(\gamma^i - {S^i \over S^+} \gamma^+ 
		\Big)\gamma_5 \psi_\alpha(0)  \nonumber \\
	& & ~~~~~~~~~~~~~~~~~~~~~~~~~~~~~~~~~~   
		 +~ h.c. \Big] |PS \rangle \, . \label{g2}
\end{eqnarray}
The operator in the above matrix element is a mixture of
operators of the transverse and plus components of bilocal axial
vector current and therefore involves twist-two and twist-three
contributions in the conventional definition. On the light-front, 
it can be expressed as 
\begin{equation}
	\overline{\psi} (\xi^-) \Big(\gamma_\bot -{S_\bot
		\over S^+} \gamma^+ \Big) \gamma_5 \psi (0)
		= {S^i\over S^+}O^s + O^m + O^{k_\bot} + O^g 
\end{equation}
which corresponds to the conventional defining twist-two operator 
$O^s$ and twist-three operators of the quark mass part $O^m$, the quark 
transverse momentum part $O^{k_\bot}$ and the quark-gluon coupling 
part $O^g$, 
\begin{eqnarray}
	&& O^s = - 2 \psi_+^\dagger (\xi^-) \gamma_5 \psi_+(0) \, , 
		\nonumber \\
	&& O^m = m \psi_+^\dagger (\xi^-) \gamma_\bot \Big({1 \over i 
		\roarrow{\partial}^+} - {1\over i \loarrow{\partial}^+} 
		\Big)\gamma_5 \psi_+(0) \, , \nonumber \\
	&& O^{k_\bot}= -\psi_+^\dagger (\xi^-)\Big(\gamma_\bot {1\over
		\roarrow{\partial}^+}{\not \! \roarrow{\partial_\bot}} 
		+ {\not \! \loarrow{\partial_\bot}}{1\over \loarrow{
		\partial}^+}\gamma_\bot \Big) \gamma_5\psi_+(0) \, ,
		\nonumber \\
	&& O^g = g\psi_+^\dagger(\xi^-) \Big ({\not \! \! A_\bot}(\xi^-)
		{1\over i\loarrow{\partial}^+}\gamma_\bot - \gamma_\bot 
		{1\over i\roarrow{\partial}^+}{\not \! \! A_\bot}(0) 
		\Big)\gamma_5 \psi_+(0) ~ \label{go}
\end{eqnarray} 
where $m$ and $g$ are the quark mass and quark-gluon coupling 
constant in QCD, and $A_\bot=A^a_{\bot}T_a$ is the transverse 
gauge field. The main concern in the previous works is which 
operator dominates $g_2$ in hard processes in terms of 
twist analysis.

However, Eqs. (\ref{g2}) and (\ref{g1}) show that $g_T$ is determined
by Lorentz invariant form factor of the bilocal axial vector 
current, $\overline{A}_{1\alpha}$, while $g_1$ is determined
by both $\overline{A}_{2\alpha}$ and  $\overline{A}_{1\alpha}$. 
According to our analysis on twist, both $\overline{A}_{1\alpha}$ 
and $\overline{A}_{2\alpha}$
are twist-two invariant matrix elements because their mass 
dimensions are zero. Although these operators are apparently so 
complicated, their contributions (in terms of matrix elements)
to $g_2$ are indeed twist-two regardless if they depend on the 
quark mass, quark transverse 
momentum and the direct quark-gluon coupling. As a matter of 
fact, it is not meaningful to separate $g_2$ into  twist-two 
and twist-three parts as in Eq.(\ref{g223}).  In our recent work 
on the validity of Wandzura-Wilczek relation, we have shown
that such a separation is strongly violated in perturbative QCD. 
The violation originates from the cancellation between the naive
twist-two and twist-three contributions due to the 
multi-parton structure of wave functions \cite{Hari97b}. 
Such a cancellation is hardly understood in the conventional twist 
expansion if one believes that the higher twist is suppressed
by an order of $1/Q$ in comparison with the lower twist. Now, we
understand that there would be no such ambiguity in the twist analysis 
contributions of $g_2$, if one used the definition of twist based on
light-front power counting and the Lorentz invariant matrix elements
of the bilocal currents (rather than the bilocal current operators
themselves).

In conclusion, the complexities of deep inelastic structure functions 
are mainly carried by the multi-parton wave functions of the hadrons 
which completely determine the Lorentz invaraint matrix element (i.e. 
the form factors) of the bilocal currents.

%%%%%%%%%%%%%%%%%%%%%%%%%%%%%%%%%%%%%%%%%%%%%%%%%%%%%%%%
\subsection{A scheme for the evaluation of soft and hard
contributions to deep inelastic structure functions}
%%%%%%%%%%%%%%%%%%%%%%%%%%%%%%%%%%%%%%%%%%%%%%%%%%%%%%%%
Up to this point, all the derivations and discussions
of the deep inelastic structure functions in the
$1/q^-$ expansion are rigorously carried out within light-front 
QCD and without recourse to perturbation theory. The remaining 
problem is how to evaluate various matrix elements of bilocal 
currents. These matrix elements contain both hard and soft 
quark and gluon dynamics. As we have analyzed in this section, 
all the hard and soft dynamics probed 
through the structure functions are completely carried by the 
target's bound state in the present formulation. This is the
main advantage of this formalism that allows us to explore the
perturbative and nonperturbative contributions to the structure
functions in the same framework. In the rest of this section, 
we shall propose a scheme for such an exploration. 

In Sec.~V.A, the hadronic bound state is  formally expressed in
terms of Fock space expansion on the light-front by Eq. (\ref{lfwf}),
and it is determined in principle by the light-front bound state
equation of Eq. (\ref{lfbe}). However, the difficulty in determining
 wave functions by solving Eq. (\ref{lfbe}) is that the QCD Hamiltonian
contains more than one energy scale. At different energy scales,
QCD Hamiltonian can exhibit different aspects of the dynamics.
Let us roughly divide the quark and gluon dynamics into
two energy domains, namely, high energy and low energy. In the 
high energy domain, the dynamics is controlled by the renormalized
QCD Hamiltonian with all the constituents carrying momenta
greater than a scale $\mu_{\rm fact}$ ($\approx 1 GeV$) 
which we call the factorization energy scale. This high 
energy QCD Hamiltonian describes all the hard dynamics of 
quarks and gluons and determines the hard contributions to the
structure functions which can be calculated in the perturbation theory. 
In the low energy domain, the effective QCD Hamiltonian is still 
unknown but such a low energy QCD Hamiltonian should fairly
determine the low energy structure of the hadrons and is
responsible for the soft contributions to the structure 
functions. 

Schematically, we may write the QCD Hamiltonian on the light-front
for DIS as
\begin{equation}
	H^{LF}_{QCD} = \left\{ \begin{array}{ll}
	 H^H_{QCD} \equiv \int_{k_{i\bot}^2 \geq \mu_{\rm fact}^2} 
		dk_i^+ d^2k_{i\bot} {\cal H}^C_{QCD} (k_i)~&~~{\rm for~hard~
		contributions} \, , \\
		~&~ \\
	 H^M_{QCD} \equiv \int dk_i^+ d^2k_{i\bot} {\cal H}^C_{QCD} 
		(k_i) &~~ {\rm for~mixed~hard~and~soft~modes}\, , \\
		~&~ \\
	 H^L_{QCD} \equiv \int_{k_{i\bot}^2 < \mu_{\rm fact}^2 } 
		dk_i^+ d^2k_{i\bot} {\cal H}^L_{QCD} (k_i) &~~{\rm for~
		soft~contributions}\, , 
			\end{array} \right.
\end{equation}
where $H^H_{QCD}$ represents the canonical light-front QCD Hamiltonian 
(with density ${\cal H}^C_{QCD}$ given in \cite{zhang93}) in which the 
transverse momenta of all the quarks and 
gluons are restricted to be $\mu^2_{\rm fact} < k_\bot^2 < Q^2$ (i.e., 
hard partons), and $H^L_{QCD}$ denotes a low energy effective 
light-front Hamiltonian in which all the constituents have the 
transverse momentum $k_\bot^2 < \mu^2_{\rm fact}$ (soft partons). 
This low energy Hamiltonian is, in principle, obtained by integrating 
out all modes with $k_\bot^2 > \mu_{\rm fact}^2$ from the canonical 
light-front QCD Hamiltonian, which leads to ${\cal H}^L_{QCD}$. In 
addition, we also introduce a 
Hamiltonian $H^M_{QCD}$ which depends only on the interaction part and
which mixes the hard and soft partons. 
Writing the light-front QCD Hamiltonian in such three parts will 
make the discussion of the perturbative and nonperturbative QCD 
contributions to DIS structure functions much more transparent, as we
will see next.

Now, the target bound state can be expressed by
\begin{eqnarray}  \label{pss}
	| PS \rangle = U_h | PS, \mu_{\rm fact}^2 \rangle \, ,
\end{eqnarray}
with
\begin{eqnarray}
	&& U_h = T^+ \exp \Bigg\{ -{i \over 2} \int_{-\infty}^0 dx^+ 
		(H^H_{QCD}+H^M_{QCD}) \Bigg\}	\, , \label{pqcd} \\ 
 && H^L_{QCD} |PS, \mu_{\rm fact}^2 \rangle = {P_\bot^2 + M^2 \over P^+}
		| PS, \mu_{\rm fact}^2 \rangle  \, .	\label{npqcd} 
\end{eqnarray}
In Eq.~(\ref{pqcd}), $H^H$ and $H^M$ contain the interaction parts only and
the mixed Hamiltonian $H^M_{QCD}$ is active only
in the extreme right of the time-ordered expansion. In other words, 
the hard and the soft dynamics in the bound states are determined 
separately by $H^H_{QCD}$ and $H^L_{QCD}$ but these two contributions 
are connected by $H^M_{QCD}$ through the time-ordered expansion of
Eq.~(\ref{pqcd}) on the state $|PS,\mu^2_{\rm fact} \rangle$ in
Eq.~(\ref{pss}), where the soft dynamics, represented by $|PS,
\mu^2_{\rm fact} \rangle$, must be solved nonperturbatively from 
Eq.~(\ref{npqcd}), and the key point to solve Eq.~(\ref{npqcd}) 
is to find the low energy effective Hamiltonian $H^L_{QCD}$. 
A practical procedure to find $H^L_{QCD}$ on the 
light-front may be the use of similarity renormalization group 
approach plus a weak-coupling treatment developed recently
\cite{Wilson94,zhang97,Perry97}. Indeed, a major effort on the study of
light-front QCD is underway at present to solve this problem \cite{review}.
%The hard dynamics can be easily carried out in time-ordered
%perturbation theory from Eq.(\ref{pqcd}). 
%A detailed calculation and analysis will be given in \cite{deep2} (also 
%see some preliminary results in \cite{Hari96}).

To see how the perturbative and nonperturbative QCD contributions
can be separately evaluated in the present formalism and how these
two contributions are connected by $H^M_{QCD}$, we substitute 
Eqs.~(\ref{pss}-\ref{npqcd}) into the expressions of structure 
functions. Denote the structure functions simply 
by $F_i \equiv: \{ F_L, F_2, g_1, g_T \}$,
\begin{eqnarray}
	F_i(x,Q^2) \sim \int d\eta e^{-i \eta x} \sum_\alpha e_\alpha^2
		\langle PS| \overline{\psi}_\alpha (\xi^-) \Gamma_i 
		\psi_\alpha (0) \pm h.c. | PS \rangle \, ,
\end{eqnarray}
where $\Gamma_i$ involves the Dirac $\gamma$-matrices [see  
Eqs. (\ref{f2+})-(\ref{gt})]. It follows that
\begin{eqnarray}
        F_i(x,Q^2) = \int d\eta e^{-i\eta x}
                && \sum_\alpha e^2_\alpha \sum_{n_1,n_2} \langle 
		P S, \mu^2_{fact}| n_1\rangle \langle n_2| PS,
		\mu^2_{fact} \rangle \nonumber \\
        && \times \langle n_1 | U_h^{-1} \Big[\overline{\psi}_\alpha
		(\xi^-) \Gamma_i \psi_\alpha (0) \pm h.c \Big]  
		U_h | n_2 \rangle \, , \label{gfock} 
\end{eqnarray}
where $|n_1\rangle, |n_2 \rangle$ are a complete set of quark and 
gluon Fock states with momentum $k_i^2 \leq \mu^2_{\rm fact}$.
This is indeed the generalized factorization theorem in the light-front
Hamiltonian formulation. The hard contribution is described by the
matrix element,
\begin{equation}  \label{hardc}
	\langle n_1 | U_h^{-1} \Big[\overline{\psi}_\alpha(\xi^-) 
	\Gamma_i \psi_\alpha (0) \pm h.c \Big] U_h | n_2 \rangle \, ,
\end{equation} 
which can be evaluated in the light-front time-ordered perturbation theory
\cite{zhang93}. The physical picture corresponds to the multi-parton
forward scattering amplitude with all the internal partons carrying a 
momentum
with the transverse component $k_{\bot}$: $\mu^2_{\rm fact} \leq k^2_\bot 
\leq Q^2 $ and the longitudinal momentum fraction $y$: $ x \leq y \leq 1$.  
The soft contribution is characterized by the overlap of the multi-parton 
wave functions in different Fock states: 
\begin{equation}  \label{softc}
	\langle P S, \mu^2_{fact}| n_1\rangle \langle n_2| PS, 
		\mu^2_{fact} \rangle \, , 
\end{equation}
which contains all the quantum correlations and interference effects 
of multi-parton (quarks and gluons) dynamics in the low energy 
domain with $k^2_\bot < \mu^2_{\rm fact}$.  Since all the internal
partons in the time-ordered expansion of $U_h$ in Eq.~(\ref{hardc})
carry momenta $\mu^2_{\rm fact} \leq k_\bot^2 \leq Q^2$, the 
mixed Hamiltonian $H^M_{QCD}$ has the contribution only in the
extreme left and extreme right of the time-ordered products. It
is this effect that connects the hard contribution of 
Eq.~(\ref{hardc}) to the soft contribution Eq.~(\ref{softc}).
We will present more detailed discussion in \cite{deep2}.

The simple parton picture in deep inelastic processes corresponds to 
the case of $|n_1\rangle =|n_2 \rangle$ in Eq. (\ref{gfock}) with only 
one parton in $|n_1 \rangle$ actively participating in the high energy
process, all others being spectators. This immediately leads to 
\begin{eqnarray}
	F_i(x,Q^2) \sim \sum_\alpha e^2_\alpha \int_x^1 dy {\cal P}_{pp',i} 
		(y,x,{Q^2 \over\mu_{\rm fact}^2}) q_{\alpha i} 
		(y,\mu_{\rm fact}^2) \, ,
\end{eqnarray}
where the hard scattering coefficient ${\cal P}_{pp',i}$ is determined by 
\begin{equation}
        {\cal P}_{pp',i}(y,x,{Q^2\over \mu_{\rm fact}^2}) \simeq \int d\eta 
		e^{-i\eta x} \langle y, k_\bot, s| U_h^{-1} \Big[ 
		\overline{\psi}_\alpha (\xi^-) \Gamma_i \psi_\alpha (0) 
		\mp h.c. \Big] U_h | y, k_\bot, s \rangle  . \label{sqm} 
\end{equation}
Here we have denoted $| y, k_\bot, s \rangle$ ($y=k^+/P^+$) as
the active parton state.  Eq.~(\ref{sqm}) means that we have 
suppressed all references to the spectators in the states $| n_1 
\rangle $. The hard scattering coefficient is directly related to the 
so-called 
the splitting function whose physical 
interpretation is the probability to find a daughter parton $p'$
in the active parent parton $p$.
%As is well-known, the moments of the splitting functions 
%determine the $Q^2$-evolution of the leading contribution to the 
%structure functions. These splitting functions can be very easily 
%evaluated in the light-front time-ordered perturbation theory 
%\cite{deep2}. 
The quantity $q_{\alpha i}(y, \mu^2_{\rm fact})$, usually 
called the parton distribution function, is given by 
\begin{equation}
	q_{\alpha i} (y, \mu^2_{\rm fact}) = \sum_n | \langle PS, 
		\mu^2_{\rm fact} | n \rangle |^2 \, ,
\end{equation}
where $n$ runs over all the Fock states containing the active 
parton with momentum fraction $y$.  
Theoretically, the parton distributions are 
determined by solving Eq. (\ref{npqcd}). Physically, they
contain only the quantum correlations of 
multi-parton dynamics but no quantum interference effects. 
Example of such distribution functions is given by 
Eqs. (\ref{f2part}-\ref{partd}) for ${F_2(x)/x}$ 
which manifestly exhibits the simple parton picture.
For detailed calculations also see Ref.  \cite{deep2}.

The above discussions indeed constitute a presentation of factorization 
scheme in the light-front Hamiltonian formulation. The leading
hard contributions to the structure functions are given by the 
the hard scattering coefficient ${\cal P}_{pp', i}(y,x,{Q^2\over 
\mu_{\rm fact}^2})$ and a complete calculation of ${\cal P}_{pp', i}$
based on the light-front time-ordered perturbative expansion of the 
multi-parton wave functions will be presented in a subsequent 
paper \cite{deep2}. The evaluation of soft contribution to 
the structure functions, given by $q_{\alpha i}(x, \mu_{\rm fact}^2)$
remains for future investigations of nonperturbative light-front 
QCD approaches to the hadronic bound states.  Other 
higher order contributions can also be systematically evaluated from 
Eqs.~(\ref{hardc}) and (\ref{softc}) \cite{deep2}.
Thus, a unified treatment of both perturbative and nonperturbative
aspects of deep inelastic structure functions in the
same framework may emerge which permits one to overcome the obstacles
in dealing with the nonperturbative QCD dynamics in OPE 
and field theoretical parton model approaches. 

%%%%%%%%%%%%%%%%%%%%%%%%%%%%%%%%%%%%%%%%%%%%%%%%%%%%%%%%%%%%%%%%%%%%%%
%\subsection{Complexities and twists of the longitudinal unpolarized
%structure function}
%%%%%%%%%%%%%%%%%%%%%%%%%%%%%%%%%%%%%%%%%%%%%%%%%%%%%%%%%%%%%%%%%%%%%%
%Based on our analysis on the twists, the only higher twist effect
%one may measure in deep inelastic processes is the longitudinal
%unpolarized structure function $F_L(x,Q^2)$ which is determined
%by the invariant matrix element $\overline{V}_{2\alpha}$. As we
%have pointed out, $\overline{V}_{2\alpha}$ is a twist-four invariant 
%matrix element. ........
%%%%%%%%%%%%%%%%%%%%%%%%%%%%%%%%%%%%%%%%%%%%%%%%%%%%%%%%%%%%%%%%%%%%%%%%%%
\section{Physical interpretation of the structure functions from sum rules}
%%%%%%%%%%%%%%%%%%%%%%%%%%%%%%%%%%%%%%%%%%%%%%%%%%%%%%%%%%%%%%%%%%%%%%%%%
In this last section, we shall explore the physical meaning of the 
deep inelastic structure functions in our framework of light-front 
QCD. The physical meaning of the structure functions can be easily  
understood from the sum rules they obey. Some of them have been
known for long time but others are new. Sum rules generally arise 
from the existence of conservation laws. First we consider the case of
sum rules in unpolarized deep inelastic scattering for which a detailed
consideration of the energy-momentum density in QCD is necessary.  
%%%%%%%%%%%%%%%%%%%%%%%%%%%%%%%%%%%%%%%%%%%%%%%%%%%%%%%%
\subsection{Energy-momentum tensor in QCD}
%%%%%%%%%%%%%%%%%%%%%%%%%%%%%%%%%%%%%%%%%%%%%%%%%%%%%%%%
%\vskip .2in

The symmetric, gauge-invariant energy-momentum tensor in QCD is
given by
\begin{eqnarray}
	\theta^{\mu \nu} = && { 1\over 2} \overline{\psi} i \big [ 
		\gamma^\mu D^\nu +\gamma^\nu D^\mu \big ] \psi 
		-F^{\mu \lambda a} F^{\nu}_{~~\lambda a} + {1 \over 4} 
		g^{\mu \nu} (F_{\lambda \sigma a } )^2 \nonumber \\
	&& -g^{\mu \nu} \overline{\psi} \left ( i \gamma^\lambda 
		D_\lambda -m  \right ) \psi.
\end{eqnarray}
The last term vanishes using the equation of motion.
{\it Formally,} we split the energy momentum tensor into a 
$``$fermionic" part $\theta^{\mu \nu}_{q}$ and a $``$gauge bosonic" 
part $ \theta^{\mu \nu}_g$:
\begin{eqnarray}
	\theta^{\mu \nu}_q = { 1 \over 2} \overline{\psi} i \Big [ 
		\gamma^\mu D^\nu +\gamma^\nu D^\mu \Big ] \psi,
\end{eqnarray}
and 
\begin{eqnarray}
	\theta^{\mu \nu}_g =  -F^{\mu \lambda a} F^{\nu}_{~~ \lambda a } + 
		{1 \over 4} g^{\mu \nu}(F_{\lambda \sigma a })^2 ,
\end{eqnarray}
with $ F^{\nu}_{~~\lambda a} = \partial^\nu A_{\lambda a} - \partial_\lambda
A^\nu_a + g f_{abc} A^\nu_b A_{\lambda c} $.
To be consistent with the study of deep inelastic structure function 
which is formulated in $A^+=0$ gauge, we shall work in the same gauge. 

We have, for the fermionic part of the longitudinal momentum density,
\begin{eqnarray}
	\theta^{++}_{q} = i \overline{\psi} \gamma^+ \partial^+ \psi. 
		\label{denp} 
\end{eqnarray}
\begin{eqnarray}
	\theta^{++}_g = - F^{+ \lambda} F^{+}_{~~\lambda} = \partial^+ A^i 
		\partial^+ A^i.
\end{eqnarray}
Thus
\begin{eqnarray}
	\theta^{++} =  i {\bar \psi} \gamma^+ \partial^+ \psi +
		\partial^+ A^i \partial^+ A^i,
\end{eqnarray}
free of interactions at the operator level itself. 
The longitudinal momentum operator
\begin{eqnarray}
	P^+ = { 1 \over 2} \int dx^- d^2 x_\perp \theta^{++}.
\end{eqnarray}

Next consider the transverse momentum density
\begin{eqnarray}
	\theta^{+i}_q = { 1 \over 2} {\bar \psi} i \Big [ \gamma^+ D^i 
		+ \gamma^i D^+ \Big ] \psi = \theta^{+i}_{q-1} + 
		\theta^{+i}_{q-2},
\end{eqnarray}
with
\begin{eqnarray}
	\theta^{+i}_{q-1} = { 1 \over 2} {\bar \psi} i \gamma^+ D^i \psi 
		~~ {\rm and} ~~ \theta^{+i}_{q-2} = { 1 \over 2} 
		{\bar \psi} i \gamma^i \partial^+ \psi. 
\end{eqnarray}
For the Hamiltonian density, the fermionic part is given by 
\begin{eqnarray}
	\theta^{+-}_q = \theta^{+-(1)}_q + \theta^{+-(2)}_q,
\end{eqnarray}
with
\begin{eqnarray}
	\theta^{+-(1)}_q = i {\psi^{+}}^\dagger \partial^- \psi^+ + 
		g {\psi^{+}}^\dagger A^- \psi^+,
\end{eqnarray}
and
\begin{eqnarray}
	\theta^{+-(2)}_q = i {\psi^{-}}^\dagger \partial^+ \psi^-.
\end{eqnarray}
Using the Dirac equation for the fermion, we find that $\theta^{+-(1)}_q=
\theta^{+-(2)}_q$. Thus we have, 
\begin{eqnarray}     
	\theta^{+-}_q &&= i \overline{\psi} \gamma^- \partial^+ \psi =
		2 i {\psi^{-}}^\dagger \partial^+ \psi^-  \label{denh} \\
	&& = 2 {\psi^{+}}^\dagger \Big [ \alpha_\perp.(i \partial_\perp 
		+ g A_\perp) + \gamma^0 m \Big ] { 1 \over i \partial^+} 
		\Big [ \alpha_\perp . (i \partial_\perp + g A_\perp)
		+ \gamma^0 m \Big ] \psi^{+} \label{thetaqf}.
\end{eqnarray}
The gauge boson part of the Hamiltonian density is more complicated
\cite{zhang93}:
\begin{eqnarray}
	\theta^{+-}_g && =  - F^{+ \lambda a} F^{-}_{~~\lambda a}+ { 1 \over 4} 
		g^{+-} (F_{\lambda \sigma a})^2 = { 1 \over 4} \Big 
		(\partial^+ A^{-a}\Big )^2 + { 1 \over 2} F^{ij a } F^{a}_{ij}
		\nonumber \\
&& = (\partial^i A_a^j)^2 + 2gf^{abc}A_a^i A_b^j \partial^i A_c^j
		  + \frac{g^2}{2}
		f^{abc} f^{ade} A_b^i A_c^j A_d^i A_e^j  \nonumber \\
	&& ~~~~~~~~~~ + 2g \partial^i A_a^i \left( \frac{1}{\partial^+}
		\right) (f^{abc} A_b^j \partial^+ A_c^j + 2 (\psi^+)^{\dagger}
		T^a \psi^+ ) \nonumber \\
	&& ~~~~~~~~~~ + g^2 \left( \frac{1}{\partial^+}
		\right) (f^{abc} A_b^i \partial^+ A_c^i + 2 (\psi^+)^{\dagger}
		T^a \psi^+ ) 
	  \left( \frac{1}{\partial^+}\right)
		(f^{ade} A_d^j \partial^+ A_e^j + 2 (\psi^+)^{\dagger} T^a
		\psi^+ )     \nonumber \\
    \label{thetagf}
%&& = \left(\partial^i A^i \right)^2 + { 1\over 2} F^{ij}F_{ij} - 
%4 g { 1\over \partial^+}
%\left ( \partial^i A^i \right ) (\psi^+)^\dagger \psi^+ + 4 g^2 \Big ({ 1
%\over \partial^+} (\psi^+)^\dagger \psi^+  \Big )^2 
\end{eqnarray}
where we have used the equation of constraint for the gauge field.     

Next, we discuss the physical interpretation of the deep inelastic
structure functions on the basis of the sum rules they obey.

%%%%%%%%%%%%%%%%%%%%%%%%%%%%%%%%%%%%%%%%%%%%%%%%%%%%%%%%
\subsection{Longitudinal momentum sum rule}
%%%%%%%%%%%%%%%%%%%%%%%%%%%%%%%%%%%%%%%%%%%%%%%%%%%%%%%%
The content of the momentum sum rule is known for long time.
For completeness, we shall rederive it in our framework.
The sum rule is simply that if we add up the longitudinal 
momentum fractions carried by all the 
quarks, antiquarks, and the gluons (alternatively by the valence 
quarks, sea quarks and the gluons) in the nucleon we should get one.
>From the expression of $F_2$ in terms of the plus component
of the bilocal current matrix element given in Eq.(\ref{f2+})
%\begin{eqnarray}
%	F_2(x) = {x \over 4 \pi P^+} \int d \eta e^{-i \eta x} \sum_\alpha
%		e^2_\alpha \langle P \mid \big [\overline{\psi}_\alpha
%		(\xi^-) \gamma^+ \psi_\alpha (0) - \overline{\psi}_\alpha(0) 
%		\gamma^+ \psi_\alpha (\xi^-) \big ] \mid P \rangle \, , 
%\end{eqnarray}
we have
\begin{eqnarray}
	\int_0^1 dx F_2(x) = \left({1 \over 2 (P^+)^2}\right) \sum_\alpha 
		e^2_\alpha \langle P \mid \theta^{++}_{F\alpha} \mid P 
		\rangle \, . \label{f2pone}
\end{eqnarray}
{\it Formally}, we can define the $``$gluon structure function" \cite{pqcd} 
\begin{eqnarray}
	F_2^G(x) = { 1 \over 4 \pi P^+} \int d\eta e^{- {i \eta x }} 
		\langle P \mid F^{+ \nu a} (\xi^-)F^{+a}_{~~\nu}(0) \mid 
		P \rangle \, ,
\end{eqnarray}
so that,
\begin{eqnarray}
	\int_0^1 dx F_2^G(x) = \left({ 1 \over 2 (P^+)^2}\right) 
		\langle P \mid \theta^{++}_G \mid P \rangle \, .
\end{eqnarray}
Only if we assume $e_\alpha=1$, one can obtain the 
momentum sum rule 
\begin{eqnarray}
	\int_0^1 dx \big [ F_2 + F_2^G \big ] =1.
\end{eqnarray} 

Similarly, from Eq.(\ref{f2i}) in terms of the transverse component of bilocal current
matrix element, we have
%\begin{eqnarray}
%	{ 1 \over 2 (P^+)^2} \langle P \mid \theta^{++} \mid P \rangle=
%	{ 1 \over 2 (P^+)^2}  \langle P \mid \Big [ \theta^{++}_F + 
%		\theta^{++}_G \Big ] \mid P \rangle = 1 \, , \label{lms}
%\end{eqnarray}  
%or
\begin{eqnarray}
	\int dx F_2(x,Q^2) 
%&=& {1\over (P^+)^2} \sum_\alpha e^2_\alpha 
%		\langle PS | \overline{\psi}(0) \gamma^+ (i
%		\stackrel{\leftrightarrow}{\partial^+}) \psi_\alpha (0) 
%		|PS \rangle \nonumber \\
%	&=&  {1\over 2(P^+)^2} \sum_\alpha e^2_\alpha 
%		\langle PS | \theta_{F\alpha}^{++} |PS \rangle \, .\\
	&=& {1\over 2P^+P^i_\bot} \sum_\alpha e^2_\alpha 
		\langle PS | \overline{\psi}(0) \gamma^+ (i
		\stackrel{\leftrightarrow}{\partial^i}) \psi_\alpha (0) 
		|PS \rangle  \nonumber \\
	&=& {1\over P^+P^i_\bot} \sum_\alpha e^2_\alpha 
		\langle PS | \theta_{F\alpha}^{+i} |PS \rangle \, ,
\label{f2tone}
\end{eqnarray}
and the momentum sum rule can also be written as
\begin{eqnarray}
	{ 1 \over 2 P^+P_\bot^i} \langle P \mid \theta^{+i} \mid P \rangle=
	{ 1 \over 2 P^+P_\bot^i}  \langle P \mid \Big [ \theta^{+i}_F + 
		\theta^{+i}_G \Big ] \mid P \rangle = 1 \,  . \label{eq1}
\end{eqnarray}  

The sum rule given in Eq.(\ref{f2pone}) 
means that $F_2$ measures the  longitudinal momentum  distribution of 
quarks inside the hadrons, as known long time ago. 
From Eqs.(\ref{f2pone}) and (\ref{f2tone}) we observe that the hadron
expectation value of the longitudinal and transverse momentum densities
gives the same information, namely, the total longitudinal momentum fraction
carried by the partons.

We note here an  apparent paradox that results when one ignores the
essential complexities carried by the state.
The operators corresponding to the transverse momentum density 
explicitly depend on the
interaction since $ D^i =\partial^i - i g A^i$ and $ \theta^{+i}_{q-2}$
depends on $\psi^-$ which in turn depends explicitly on the interaction.
Since we know that $P^i$ is a kinematical operator this appears puzzling at
first sight. Thus we expect the apparent dependence of 
$P^i$ on the interaction to be spurious. However this cannot be  
demonstrated at the level of operators alone. But this is not a
serious problem since what really matters are the matrix elements. 

Indeed, our demonstration \cite{Hari97} that the matrix element 
of the transverse component of the vector bilocal has the same parton 
interpretation as that of the plus component and hence the apparent 
interaction dependence in the former is completely spurious in turns 
directly tells us that the interaction dependence of the operator 
$\theta^{+i}_{q-2}$ is completely spurious. In that case an explicit 
evaluation of off-diagonal matrix elements in the Fock space expansion
of states is involved.  Similarly, an explicit 
demonstration shows that $\theta^{+i}_{q-1}$ has no interaction 
dependence at the level of matrix element, namely
\begin{eqnarray}
	\theta^{+i}_{q-1} = { 1 \over 2} i {\bar \psi} i \gamma^+ 
		\partial^i \psi + g (\psi^+)^\dagger A^i \psi^+,
\end{eqnarray}
but the matrix elements of the second term vanishes. Then, at the 
level of matrix elements
\begin{eqnarray}
	\theta^{+i}_{q} = i {\bar \psi} \gamma^i \partial^+ \psi.
\end{eqnarray}
%{\bf Is it obvious?}. Then we have a consistent picture for the transverse
%component. 
This demonstration clearly shows that drawing conclusions by
looking at the operator structure is quite misleading in the case of
operators that are twist three in the conventional definition.

Since $F_2$ involves quark charges in specific combinations, 
it does not give the direct test of the above momentum sum rule. To
test the sum rule experimentally, one can combine the 
data for both the  electron-proton and electron-neutron 
deep-inelastic scatterings and assume that the sea is flavor symmetric, 
then
\begin{eqnarray}
 	\int dx \Big[ F_2^{ep}(x) + F_2^{en}(x) \Big] &=& {5\over 9}
		{1\over (P^+)^2} \sum_\alpha \langle PS 
		| \theta_{F\alpha}^{++} |PS \rangle \nonumber \\
	&=& {5\over 9} {1\over P^+P^i_\bot} \sum_\alpha  
		\langle PS | \theta_{F\alpha}^{+i} |PS \rangle \, .
\end{eqnarray}
This shows that ${9\over 5} \int dx \Big[ F_2^{ep}(x) + F_2^{en}(x) 
\Big]$ is the total longitudinal momentum fraction carried by the all the 
quarks in proton and neutron.
If the quarks carry all the momentum, then we expect that
\begin{equation}
	\int dx \Big[ F_2^{ep}(x) + F_2^{en}(x) \Big] = {5\over 9} \, .
\end{equation}
Experimental data shows that the above integral is $0.28$. In other
words, as is well-known, there are half of the momentum in hadrons are carried by 
gluons or the see quarks if the sea is not flavor symmetric. 
 
%%%%%%%%%%%%%%%%%%%%%%%%%%%%%%%%%%%%%%%%%%%%%%%%%%%%%%%%%%%%%%%%
\subsection{ Sum rule for $g_1$}
%%%%%%%%%%%%%%%%%%%%%%%%%%%%%%%%%%%%%%%%%%%%%%%%%%%%%%%%%%%%%%%%%
Now we consider the sum rule for $g_1$ and 
its physical interpretation. Integrating $g_1$ over $x$,
we simply have
\begin{equation}
	\Gamma_1(Q^2) = \int dx g_1(x,Q^2) = {1\over 2S^+} \sum_\alpha 
		e^2_\alpha \langle PS | \overline{\psi}_\alpha(0) 
		\gamma^+ \gamma_5 \psi_\alpha(0) | PS \rangle 
\end{equation}
Note that 
\begin{equation}  \label{pc1}
	\langle PS | \overline{\psi}_\alpha(0) \gamma^+ \gamma_5
	 \psi_\alpha(0) | PS \rangle = \Delta q^{GI}_\alpha (Q^2)S^+
\end{equation}
%or
%\begin{equation}  \label{pc2}
%	\langle PS | \overline{\psi}_\alpha(0) \gamma^+ \gamma_5
%	  \psi_\alpha(0) | PS \rangle = \Big(\Delta q^{CI}_\alpha (Q^2)
%		- {\alpha_s \over 2\pi} \Delta G (Q^2)\Big) S^+
%\end{equation}
where $\Delta q_\alpha^{GI}$ is the distribution function of 
chirality carried by all the quarks in the  longitudinally polarized 
target and renormalized in the gauge invariant scheme.
The above form can also be directly obtained from Eq.
(\ref{bacff}). Then, we have 
\begin{equation} \label{g1sr}
	\Gamma_1(Q^2) = {1\over 2} \sum_\alpha e^2_\alpha
		\Delta q_\alpha^{GI} (Q^2) \, .
%= {1\over 2} \sum_\alpha
%		e^2_\alpha \Big( \Delta q_\alpha^{CI}(Q^2) - 
%		{\alpha_s \over 2\pi} \Delta G (Q^2) \Big) \, .
\end{equation}
If one uses the chiral invariant renormalization scheme, the
first moment of $g_1$ also exhibits the anomaly contribution
\cite{Cheng96}. We will discuss how this property manifests
in the light-front Hamiltonian formulation in a separate
publication.
 
It is clear now that $g_1$ describes the distribution of 
chirality carried by the quarks inside the target (proton or neutron). 
Note the first moment $\Gamma_1$ is usually called the proton's spin 
structure function. But one must also be aware that on the light-front, 
the plus component of axial current is the same as the third component
of the quark helicity operator density on the light-front. Therefore, its
expectation value is the same as the third component of the spin on the 
light-front (i.e., light-front helicity), which is not the same
as the $z$-component of intrinsic spin defined in the rest frame of
the equal-time coordinates. On the light-front, as is well-known,
there is a very complicated relation between the light-front helicity
and the intrinsic spin in the rest frame. This relation depends on the 
interactions in the fundamental theory. At present, one only knows
the exact relation for free theory \cite{melosh}. In other words, 
$g_1$ does not really measure the spin of proton. Simply calling 
Eq. (\ref{g1sr}) a spin sum rule is misleading. 

%%%%%%%%%%%%%%%%%%%%%%%%%%%%%%%%%%%%%%%%%%%%%%%%%%%
\subsection{Light-front helicity sum rule}
%%%%%%%%%%%%%%%%%%%%%%%%%%%%%%%%%%%%%%%%%%%%%%%%%%%
For the fermions, the intrinsic light-front helicity distribution function 
is given by
\begin{eqnarray}
\Delta q(x,Q^2) = { 1 \over 8 \pi S^+} \int d \eta e^{ - i \eta x} 
\langle PS \mid \Big [ {\overline \psi} (\xi^- ) \gamma^+ \Sigma^3 \psi(0) +
h.c \Big ] \mid PS \rangle
\end{eqnarray}
where $ \Sigma^3 = i \gamma^1 \gamma^2$. This is the same as the chirality
distribution function $g_1$.

We define the orbital helicity distribution for the fermion
\begin{eqnarray}
\Delta q_L(x, Q^2) = { 1 \over 4 \pi P^+} \int d \eta e^{ -i  \eta x} 
\langle PS \mid \Big [ {\overline \psi}(\xi^-) \gamma^+ i (x^1 \partial^2
- x^2 \partial^1) \psi(0) + h.c.\Big ] \mid PS \rangle.
\end{eqnarray}
For the gluon, the intrinsic light-front helicity distribution is defined
\cite{jgd} as
\begin{eqnarray}
\Delta g(x,Q^2) = -{ i \over 4 \pi (P^+)^2 x} \int d \eta e^{ - i \eta x}
\langle PS \mid F^{+ \alpha} (\xi^-) {\tilde F}^+_{~~\alpha(0)} \mid PS \rangle.
\end{eqnarray}
The dual tensor 
\begin{eqnarray}
{\tilde F^{\mu \nu}} = { 1 \over 2} \epsilon^{\mu \nu \rho \sigma} F_{\rho
\sigma} ~~~~
{\rm with} ~~~~ \epsilon^{+1-2} = 2.
\end{eqnarray}
We also define the light-front orbital helicity distribution for the gluon as
\begin{eqnarray}
	\Delta g_L(x,Q^2) &=& -{1  \over 4 \pi P^+} \int d \eta 
		e^{ - i \eta x} \langle PS \mid \Big [ x^1 F^{+ \alpha}(\xi^-) 
		\partial^2 A_\alpha(0) \nonumber \\
	& & ~~~~~~~~~~~~~~~~~~~~~~~ - x^2 F^{+ \alpha} (\xi^-) \partial^1 
		A_\alpha(0) \Big ] \mid PS \rangle. 
\end{eqnarray}
Note that all the above distribution functions are defined in the
light-front gauge $A^+=0$. 

The light-front helicity operator is given by 
\begin{eqnarray}
J^3 = {1  \over 2} \int dx^- d^2 x^\perp 
\Big [ x^1 \theta^{+2} - x^2 \theta^{+1} \Big
]
\end{eqnarray}
where $\theta^{\mu \nu}$ is the symmetric energy momentum tensor.
Explicitly,
the fermion orbital helicity operator
\begin{eqnarray}
J^3_{q(o)}=i \int dx^- d^2 x^\perp {\psi^+}^{\dagger} \Big [ x^1 \partial^2 -
x^2 \partial^1 \Big ] \psi^+,
\end{eqnarray}
and the fermion intrinsic helicity operator 
\begin{eqnarray}
J^3_{q(i)}=  { 1 \over 2} \int dx^- d^2 x^\perp {\psi^+}^\dagger 
\Sigma^3 \psi^+.
\end{eqnarray}
The gluon orbital helicity operator
\begin{eqnarray}
J^3_{g(o)} = { 1 \over 2} \int dx^- d^2 x^\perp \Big [ x^1 (\partial^+ A^1
\partial^2 A^1+ \partial^+ A^2 \partial^2 A^2)- x^2 ( \partial^+ A^1
\partial^1 A^1 + \partial^+A^2 \partial^1 A^2) \Big ]
\end{eqnarray}
and the gluon intrinsic helicity operator 
\begin{eqnarray}
J^3_{g(i)} = { 1 \over 2 } \int dx^- d^2 x^\perp \Big [ A^1 \partial^+ A^2 -
A^2 \partial^+ A^1 \Big ].
\end{eqnarray}

The helicity sum rule for the nucleon target implies
\begin{eqnarray}
{ 1 \over {\cal N}} \langle PS \mid \Big [ J^3_{q(i)} + J^3_{q(o)} +
J^3_{g(i)} + J^3_{g(o)} \Big ] \mid PS \rangle = \pm { 1 \over 2},
\end{eqnarray}
where $ {\cal N} = 2 ( 2 \pi)^3 P^+\delta^3(0)$.
Thus we arrive at the sum rule obeyed by the helicity distribution functions
\begin{eqnarray}
\int_0^1 dx \Big [ \Delta q(x,Q^2) + \Delta q_L(x,Q^2) 
+ \Delta g(x,Q^2) + \Delta g_L(x,Q^2) \Big ] = \pm {1 \over 2}
\end{eqnarray}
as a result of light-front helicity conservation.
 
%Experimentally, $\Gamma_1^p = 0.126 \pm 0.010$. If only the valence
%quarks contribute to $\Gamma_1$, then, 
%\begin{equation}
%	\Gamma_1 (Q^2) = {1\over 2} \Big({4\over 9}\Delta u(Q^2)
%		+ {1\over 9} \Delta d(Q^2) + {1\over 9} \Delta
%		s(Q^2) \Big) 
%\end{equation}
%.....
%%%%%%%%%%%%%%%%%%%%%%%%%%%%%%%%%%%%%%%%%%%%%%%%
\subsection{Sum rule for $g_T$ and the Burkhardt-Cottingham sum rule}
From Eq.(\ref{g10}) it follows that
\begin{eqnarray}
\int_0^1 dx g_1(x,Q^2) = { 1 \over 16 \pi} \int_{- \infty}^{+ \infty} dx 
\int d \eta e ^{ - \eta x} \sum_\alpha e_\alpha^2 \Big (
{\overline A}_{1 \alpha} + { 1 \over 2} P^+ \xi^- 
{\overline A}_{2 \alpha} \Big ).
\end{eqnarray}
Provided  the bilocal form factor ${\overline A}_{2 \alpha} $ does not have
pathological behavior as $ \xi^- \rightarrow 0$, we have,
\begin{eqnarray}
\int_0^1 dx g_1(x,Q^2) = { 1 \over 8}{\overline A}_{1 \alpha} (0).
\label{gisr}
\end{eqnarray}
Also we have, from Eq.(\ref{gt0}),
\begin{eqnarray}
\int_0^1 dx g_T(x,Q^2) = { 1 \over 8}{\overline A}_{1 \alpha} (0).
\label{gtsr}
\end{eqnarray}
Since, $g_T = g_1+g_2$, it follows from Eqs.(\ref{gisr}) and (\ref{gtsr}) that
\begin{eqnarray}
\int_0^1 dx g_2(x, Q^2) =0
\end{eqnarray}
which is the Burkhardt-Cottingham sum rule.
 
Recently, in the literature, there have been discussions about 
the validity of BC sum rule in perturbative  QCD\cite{pqcdg2}. 
Here we have shown the validity of the sum rule exactly up
to the leading contribution in the $1/q^-$ expansion without 
recourse to perturbation theory.

Obviously, the BC sum rule does not provide us any intuition about 
the physical picture of $g_2$. Indeed, as we have pointed out  
the physical picture for $g_2$ is not clear, since experimentally
one directly measures $g_1$ and $g_T$ when the target is polarized
in the longitudinal and transverse directions, respectively. The 
transverse polarized structure function is $g_T$ rather than $g_2$. 
Eqs. (\ref{gisr}) and (\ref{gtsr}) indicates that by averaging 
over $x$, the longitudinal and the transverse structure functions 
give the same result. This can be again regarded as a consequence of 
rotational symmetry. However, this does not implies that $g_1(x)$ and 
$g_T(x)$ are the same.

To see clearly the intrinsic physical picture of $g_T(x,Q^2)$,
let us consider the target state being transversely polarized in 
the $x$-direction without loss of generality. Then we can simply 
express $|PS\rangle$ 
as a combination of the helicity up and down states: $|P S_\bot^x
\rangle ={1\over \sqrt{2}}(|P\uparrow\rangle \pm |P\downarrow
\rangle)$ for $S^x=\pm M$. It is easy to show that {\it
$g_T$ measures the helicity flip processes on the light-front}
\cite{zhang96},
\begin{eqnarray}
        g_T(x,Q^2) &=& {1\over 8\pi M} \int_{-\infty}^{\infty}
                d\eta e^{-i\eta x} {1\over 2} \sum_\lambda \langle P 
		\lambda |\overline{\psi}_\alpha(\xi^-) \Big(\gamma^i 
		- {P^i\over P^+} \gamma^+ \Big)\gamma_5 \psi_\alpha(0) 
		\nonumber \\
              & & ~~~~~~~~~~~~~~~~~~~~~~~~~~~~~~~~~~~~~  + h.c | 
		P\!-\!\lambda \rangle .  \label{helib} 
\end{eqnarray}                 
Not that the quantity $g_2$ is purely introduced
in the Lorentz decomposition of the hadronic tensor $W^{\mu \nu}$
for historical reasons and has no clear physical interpretation. 
Only $g_1$ and $g_T$ have the clear physical picture: $g_1$ 
measures the parton helicity distribution and $g_T$ measures 
the parton helicity flip effect, which is equivalent 
to the measurement of the effects of chiral symmetry breaking and 
therefore it involves more complicated intrinsic dynamics of quarks and 
gluons. A possible relation between $g_T$ and dynamical chiral 
symmetry breaking is explored in  \cite{zhang96}, and more detailed 
theoretical and experimental investigations remain to be further
carried out.

%%%%%%%%%%%%%%%%%%%%%%%%%%%%%%%%%%%%%%%%%%%%%%%%%%%
\subsection{Sum rule for twist four part of $F_L$}
%%%%%%%%%%%%%%%%%%%%%%%%%%%%%%%%%%%%%%%%%%%%%%%%%%%
>From Eqs. (\ref{f2+}) and (\ref{fl}) it follows that $ F_{2}(-x) = F_{2}(x)$
 and  $F_{L}^{\tau=4}(-x) = 
-F^{\tau=4}_{L}(x) $. Consider the integral
\begin{eqnarray}
	\int_{- \infty}^{+ \infty}dx {F^{\tau=4}_{L}(x) \over x}&& = 
		2 \int_{0}^{ \infty}dx {F^{\tau=4}_{L}(x)
		\over x} \nonumber \\
	&&= {P^+ \over \pi Q^2 } \int_{- \infty}^{+ \infty} x dx 
		 \int d\eta e^{-i \eta x } \sum_\alpha
		e^2_\alpha \langle P \mid \overline{\psi}_\alpha (\xi^-)
		\Big(\gamma^- \nonumber \\
	&& ~~~~~~~~~~~~~~~~~ - {(P_\perp)^2 \over (P^+)^2} \Big)
		\psi_\alpha(0) - h.c. \mid P \rangle \Bigg\}
\end{eqnarray}
where in the first equality, the symmetry property of $F^{\tau=4}_{L}$ 
has been used.
Interchanging the orders of $x$ and $y^-$ integrations and carrying out the
integrations explicitly, we arrive at \cite{hari97c}
\begin{eqnarray}
	\int_0^1 dx  {F_{L}^{\tau=4}(x,Q^2) \over x} = { 2 \over Q^2} 
		\sum_\alpha e^2_\alpha \Big [ \langle P \mid \overline{\psi}
		_\alpha(0) i \Big(\gamma^- \partial^+ - {(P_\perp)^2 
		\over (P^+)^2} \gamma^+ \partial^+ \Big) \psi_\alpha
		{(0)} \mid P \rangle \Big ]. 
\end{eqnarray}
Identifying $ i \overline{\psi} \gamma^- \partial^+ \psi = \theta^{+-}_q$, 
the fermionic part of the light-front QCD Hamiltonian density and 
$i \overline{\psi} \gamma^+ \partial^+  \psi = \theta^{++}_q$, the fermionic
part of the light-front QCD longitudinal momentum density, 
(see Eqs. (\ref{denp}) and (\ref{denh}) above), we arrive at 
the interesting relation:
\begin{eqnarray}
	\int_{0}^{1} dx {F_{L}^{\tau=4}(x,Q^2) \over x} = { 2 \over Q^2} 
		\sum_\alpha e^2_\alpha \Big [ \langle P \mid \Big( 
		\theta_{q\alpha}^{+-}(0) - {(P_\perp)^2 \over (P^+)^2} 
 		\theta^{++}_{q\alpha}(0) \mid P \rangle \Big ], \label{flsr1}
\end{eqnarray}    
We have used the fact that the physical structure function vanishes for 
$x >1$. We observe that the integral of ${F^{\tau=4}_{L(q)} \over x}$ is 
related to the hadron matrix element of the fermionic parts of the 
light-front {\it Hamiltonian density}. The above relation makes
manifest the non-perturbative nature of the twist-four part of the
longitudinal structure function.

The fermionic operator matrix elements appearing in Eq. (\ref{flsr1}) 
changes with $Q^2$ as a result of the mixing of quark and gluon operators in
QCD under renormalization.    
Next we analyze this problem of operator mixing and derive a new 
sum rule at the twist four level arising as a result of the conservation of
energy-momentum tensor. 

We define the twist four longitudinal gluon structure function
\begin{eqnarray}
	F_{L(g)}^{\tau=4}(x) && = { 1 \over Q^2} {x P^+ \over 2 \pi} 
		\int d\xi^- ~e^{-{i \eta x}} \nonumber \\
&& ~~~~~\Big [ \langle P \mid (-) F^{+ \lambda a}(\xi^-) F^-_{\lambda a}(0) + 
	{ 1 \over 4} g^{+-} F^{\lambda \sigma a} (\xi^-) F_{\lambda \sigma a}
	(0) \mid P \rangle \nonumber \\
&& ~~~~~~~ - {(P^\perp)^2 \over (P^+)^2} \langle P \mid F^{+ \lambda a}(\xi^-) 
F^+_{\lambda a}(0) \mid P \rangle \Big ].
\end{eqnarray}
Then if we assume $e_\alpha=1$, we have,
\begin{eqnarray}
	\int_0^1 { dx \over x} \Big [ F_{L}^{\tau=4} + F_{L(g)}^{\tau=4} 
		\Big ] = { 2 \over Q^2} \Big [ \langle P \mid \theta^{+-}
		(0) \mid P \rangle - {(P^\perp)^2 \over (P^+)^2} \langle P 
		\mid \theta^{++}(0) \mid P \rangle \Big ] \, . \label{flsr2}
\end{eqnarray}
But,
\begin{eqnarray}
	\langle P \mid \theta^{+-}(0) \mid P \rangle = 2 P^+ P^- = 2 (M^2 +
	(P^\perp)^2)~~ {\rm and} ~~ \langle P \mid \theta^{++}(0) \mid P 
		\rangle = 2 (P^+)^2 \, ,
\end{eqnarray}
where $M$ is the invariant mass of the hadron. Thus we arrive at a new 
sum rule for the twist four part of the longitudinal structure function
\cite{hari97c}
\begin{eqnarray}
	\int_0^1 { dx \over x} \Big(F_L^{\tau=4} + F_{L(g)}^{\tau=4}\Big) 
		= 4 {M^2 \over Q^2} \, . \label{flsr}   
\end{eqnarray}

\section{Conclusion}

In this paper, beginning with an inverse power expansion of the 
light-front energy of the probe in the framework of light-front 
QCD, we have arrived at the most general expression for the leading
contributions to deep inelastic structure functions as the Fourier 
transform of the matrix element of different components of bilocal 
vector and axial vector currents. Although some of the expressions 
are already known, others are either completely new,
such as the expression for $F_2$ in terms of the transverse component
of bilocal current matrix element and the expression for $F_L$, 
or generalizations of  earlier results
in some specific Lorentz frame to an arbitrary Lorentz frame, e.g.
the expressions for $g_T$ (or $g_2$). 

We have also provided a 
consistent and Lorentz invariant definition of twist based
on light-front power counting. Using the light-front power counting, 
we found that quark-gluon coupling operators contributing
to $g_2$ are indeed twist-two when we look at their 
Lorentz invariant matrix elements. Only $F_L$ has the leading 
twist-four contribution. No twist-three Lorentz invariant 
matrix element emerges in the leading contributions
of the inclusive deep inelastic 
lepton-nucleon scatterings. The light-front power counting 
\cite{Wilson94} naturally separates the space-time dimensions into 
two different scales measured in deep inelastic processes, the 
transverse scale and the longitudinal scale. Only the transverse 
scale has the mass dimension which determines the $1/Q^2$ suppressions.
Therefore the light-front mass dimension seems to be more meaningful
than the concept of twist in the analysis of the $1/Q^2$ suppressions.
Light-front power counting analysis may eliminate some confusions
in the discussion of higher-twist contributions to structure functions, 
especially, the transverse polarized structure function.

We have also derived new sum rules for
$g_T$ and $F_L$, which provide the  physical picture of
these structure functions. An important feature of the present
formulation of deep inelastic processes is the
fact that we have unified the treatment of soft and hard 
contributions to the structure functions in terms of multi-parton
wave functions. The hard contributions can be easily calculated
from light-front time-ordered perturbative expansion  
in high energy QCD  \cite{deep2,zhang93,Hari96}, while the soft
contributions can be evaluated by the multi-parton wave functions below
the factorization scale ($ \approx 1$ GeV), by solving the 
light-front bound state equation based on the recently developed 
nonperturbative renormalization group approach 
\cite{Wilson94,zhang97,Perry97} or other approaches
on light-front QCD \cite{review}.
Further investigations along this direction are in progress and will
be published in forthcoming papers.  

\acknowledgments
We acknowledge enlightening discussions 
with Stan Brodsky and James Vary.
WMZ also thanks H. Y. Cheng, C. Y. Cheung
and H. L. Yu for fruitful discussions. 
This work is partially supported by NSC86-2816-M001-009R-L and
NSC86-2112-M001-020 (WMZ).

\end{document}